\documentclass[]{aastex631}

\usepackage{CJKutf8}

\def \NAU {Department of Astronomy and Planetary Science, Northern
  Arizona University,\\PO Box 6010, Flagstaff, AZ 86011, USA}
\def \UMPhysics {Department of Physics, University of Michigan,\\ Ann
  Arbor, MI 48109, USA}
\def \UMAstronomy {Department of Astronomy, University of
  Michigan,\\ Ann Arbor, MI 48109, USA}
\def \UW {DiRAC Institute and the Department of Astronomy, University
  of Washington, Seattle, USA}
\def \uchile {Departamento de Astronomía, Universidad de
  Chile,\\ Camino del Observatorio 1515, Las Condes, Santiago, Chile}
\def \cfa {Harvard-Smithsonian Center for Astrophysics,\\ 60 Garden
  St., MS 51, Cambridge, MA 02138, USA}
\def \byu {Department of Physics and Astronomy, Brigham Young
  University, Provo, UT 84602, USA}
\def \apl {Applied Physics Lab, Johns Hopkins University,\\ 11100
  Johns Hopkins Road, Laurel, Maryland 20723, USA}
\def \ucla {Department of Earth, Planetary and Space Sciences,
  University of California Los Angeles, 595 Charles E. Young Dr. East,
  Los Angeles, CA 90095, USA}
\def \carnegie {Earth and Planets Laboratory, Carnegie Institution for
  Science, Washington, DC 20015}
\def \stgallen {School of Computer Science, University of
  St. Gallen,\\ Rosenbergstrasse 30, CH-9000 St. Gallen, Switzerland}

\shorttitle{DEEP II. Observational Strategy and Design}
\shortauthors{Trujillo, C. A. et al.}

\begin{document}

\title{The DECam Ecliptic Exploration Project (DEEP) II. Observational Strategy and Design}

\correspondingauthor{Chadwick A. Trujillo}
\email{chad.trujillo@nau.edu}

\author[0000-0001-9859-0894]{Chadwick A. Trujillo}
\affiliation{\NAU}

\author[0000-0002-5211-0020]{Cesar Fuentes}
\affiliation{\uchile}

\author[0000-0001-6942-2736]{David~W.~Gerdes}
\affiliation{\UMPhysics}
\affiliation{\UMAstronomy}

\author[0000-0002-2486-1118]{Larissa Markwardt}
\affiliation{\UMPhysics}

\author[0000-0003-3145-8682]{Scott S. Sheppard}
\affiliation{\carnegie}

\author[0000-0001-6350-807X]{Ryder Strauss}
\affiliation{\NAU}

\author[0000-0001-7335-1715]{Colin Orion Chandler}
\affiliation{\UW}
\affiliation{LSST Interdisciplinary Network for Collaboration and
  Computing, 933 N. Cherry Avenue, Tucson AZ 85721}
\affiliation{\NAU}

\author[0000-0001-5750-4953]{William J. Oldroyd}
\affiliation{\NAU}

\author[0000-0003-4580-3790]{David E. Trilling}
\affiliation{\NAU}

\author[0000-0001-7737-6784]{Hsing~Wen~Lin (\begin{CJK*}{UTF8}{gbsn}
                                林省文 \end{CJK*})
}
\affiliation{\UMPhysics}

\author[0000-0002-8167-1767]{Fred C.~Adams}
\affiliation{\UMPhysics}
\affiliation{\UMAstronomy}

\author[0000-0003-0743-9422]{Pedro H. Bernardinelli}
\affiliation{\UW}

\author[0000-0001-8550-6788]{Matthew J. Holman}
\affiliation{\cfa}

\author[0000-0003-1996-9252]{Mario Juri\'c}
\affiliation{\UW}

\author{Andrew McNeill}
\affiliation{\NAU}
\affiliation{Department of Physics, Lehigh University, 16 Memorial
Drive East, Bethlehem, PA, 18015, USA}

\author[0000-0002-7817-3388]{Michael Mommert}
\affiliation{\stgallen}

\author[0000-0003-4827-5049]{Kevin J. Napier}
\affiliation{\UMPhysics}

\author[0000-0001-5133-6303]{Matthew J. Payne}
\affiliation{\cfa}

\author[0000-0003-1080-9770]{Darin Ragozzine}
\affiliation{\byu}

\author[0000-0002-9939-9976]{Andrew S. Rivkin}
\affiliation{\apl}

\author{Hilke Schlichting}
\affiliation{\ucla}

\author[0000-0002-7895-4344]{Hayden Smotherman}
\affiliation{\UW}





\begin{abstract}

We present the DECam Ecliptic Exploration Project (DEEP) survey
strategy including observing cadence for orbit determination, exposure
times, field pointings and filter choices. The overall goal of the
survey is to discover and characterize the orbits of a few thousand
Trans-Neptunian Objects (TNOs) using the Dark Energy Camera (DECam) on
the Cerro Tololo Inter-American Observatory (CTIO) Blanco 4 meter
telescope. The experiment is designed to collect a very deep series of
exposures totaling a few hours on sky for each of several 2.7 square
degree DECam fields-of-view to achieve magnitude $\sim 26.2$ using a
wide $V\!R$ filter which encompasses both the $V$ and $R$
bandpasses. In the first year, several nights were combined to achieve
a sky area of about 34 square degrees. In subsequent years, the fields
have been re-visited to allow TNOs to be tracked for orbit
determination. When complete, DEEP will be the largest survey of the
outer solar system ever undertaken in terms of newly discovered object
numbers, and the most prolific at producing multi-year orbital
information for the population of minor planets beyond Neptune at 30
au.

\end{abstract}


\section{Introduction}
\label{introduction}

For decades, the properties of the Kuiper Belt and Trans-Neptunian
region have been, and still remain to this day, observationally
limited. The only way to improve this situation is by discovering more
objects and characterizing their orbital and physical properties. As
of June 1, 2023, there are 2,429 multi-opposition Trans-Neptunian
Objects (TNOs) listed by the Minor Planet Center. Survey completion
rates for the brightest objects are higher than for other TNO
populations, with perhaps $\gtrsim 75\%$ of the dwarf planet-sized
objects known in the Northern hemisphere \citep{2014AJ....147....2S}
and a similar efficiency in the Southern hemisphere
\citep{2011AJ....142...98S,2012AJ....144..140R} out to heliocentric
distances of $\sim 100$ au. These large bodies have diameters $D
\gtrsim 450$ km \citep{2008Icar..195..851T,2020tnss.book..395P} and
typically are brighter than red magnitude $m_R \lesssim 20$ interior
to heliocentric distances of $R \lesssim 40$ au. However, since there
are far more smaller TNOs ($D \lesssim 100$ km, $m_R \gtrsim 23$) than
large TNOs, the known sample of $D \sim 100$ km TNOs is a small
fraction of the inferred population. Estimates of the small population
vary, but recent ground-based surveys with hundred of detections find
the number density of TNOs near the ecliptic at $m_R \simeq 24.5$
(about $D \lesssim 25$ km for albedos of 15\%) to be more than 5
objects per square degree \citep{2018ApJS..236...18B} suggesting that
there are tens of thousands of such bodies which remain undiscovered,
a population size that has been refined, but not changed in overall
magnitude in decades \citep{2001AJ....122..457T}. It is clear that our
knowledge of the number of TNOs is far from complete --- likely more
incomplete than any other stable solar system population except for
the extremely distant Extreme Trans-Neptunian Objects (ETNOs) beyond
50 au \citep{2019AJ....157..139S}.

It is this main deficiency of our knowledge of the TNOs that was the
impetus for our project. The main science goals and how they impact
our observational design are discussed in more detail in Section
\ref{principles}. Although these science goals have been previously
studied by other surveys, none have done so in as great detail as in
this survey. In the Dark Energy Camera (DECam) Ecliptic Exploration
Project (DEEP), we expect to discover a few thousand new TNOs and
trace their orbits over several years. In this work, we discuss the
observational choices that were made to achieve this goal and to
achieve the scientific goals of our project. Other works discuss the
methods used to discover new bodies
\citep{2022AAS...24022705G,2019arXiv190108549J} and investigations of
other populations such as the main belt asteroids
\citep{2021DPS....5320204T}.

We expect that our survey, when complete, will be an improvement on
prior surveys in terms of discovery statistics by factors of
a few. Prior works can largely be classified into three categories:
(1) very wide-area ($> 1,000$ square degrees) but shallow ($m_R < 21$)
surveys which are designed to maximize the discovery rate of dwarf
planet sized objects
\citep{2003EMP...92...99T,2009ApJ...694L..45S,2010ApJ...720.1691S,2011AJ....142...98S,2012AJ....144..140R},
(2) moderately wide-area surveys ($\sim 100$ square degrees)
investigating somewhat fainter magnitudes ($m_R \sim 24$)
\citep{2001AJ....122..457T,2005AJ....129.1117E,2011AJ....142..131P,2018ApJS..236...18B,2019AJ....157..139S,2020ApJS..247...32B,2022ApJS..258...41B}
and (3) very deep ($m_R > 24.5$) but small-area surveys ($< 50$ sq
deg) probing the faintest TNOs
\citep{1998AJ....116.2042G,1999AJ....118.1411C,2004AJ....128.1364B,2010Icar..209..766P}. It
is beyond the scope of this work to summarize these (and many other)
surveys --- for a more comprehensive work on these surveys and many
other aspects of the TNOs, see \cite{2020tnss.book.....P}.

One of our primary survey goals was to maximize total number of
objects discovered with orbital information. This simple choice
created several constraints on our survey in terms of optimization:
(1) a survey duration of 1--3 years was required to understand TNO
orbits, which tend to have orbital periods of order $\sim 200$ years
for bodies at 50 au and nearer, (2) since the size distribution of the
TNOs strongly favors faint bodies, a large telescope ($\gtrsim 4$ m)
was needed to reach magnitudes of $m_R > 24$, and (3) since total
number of bodies was important, and the TNOs are known to have an
ecliptic plane density of about 1 body per square degree for magnitude
$m_R \sim 24$, our telescope needed to have a wide field of view given
its size. With these parameters in mind, our instrument of choice was
the DECam instrument mounted at prime focus on the V\'{i}ctor
M. Blanco 4 m telescope atop Cerro Tololo, Chile. These choices also
affected our observing strategy, which was to maximize survey depth,
meaning that we would observe for typically $\sim 4$ hours on sky for
any given field.

The DECam instrument on the Blanco 4 m is an excellent instrument for
this study due to its large field of view ($\sim 2.7$ square degrees)
on a large telescope ($D \sim 4$ m) with $\sim 1\arcsec$
full-width-at-half-maximum seeing at a site relatively protected from
light pollution, which compares well with other survey instruments in
terms of lightgrasp \citep{2008ssbn.book..573T}. Our overall survey
expectations (discussed in more detail in Section~\ref{observing}) are
that we should be able to probe to limiting magnitude $m_R \sim 26.2$
over 32 -- 108 square degrees over a few years, which will generate
the discovery of a few thousand new TNOs and allow orbital properties
to be constrained for many of them.

This total number of objects should significantly exceed prior works,
which for any given work have discovered $< 1,000$ total objects and
in many cases far fewer. Our survey is complementary to the expected
data characteristics to be produced by the Vera C. Rubin Observatory
Legacy Survey of Space and Time (LSST). The LSST has an exposure time
of about 30 s on a 6.7 m effective aperture telescope, resulting in a
survey depth of about $m_R \sim 24.5$
\citep{2009arXiv0912.0201L,2021BAAS...53d.236V}. Since it will survey
the entire sky available from Cerro Pach\'{o}n, which includes most of
the ecliptic, it should find some $\sim 40,000$ TNOs brighter than
$m_R \sim 24.5$. Although the total number of TNOs discovered by the
LSST, which is a billion dollar project planned to span about a decade
of operations, will far exceed our survey, our survey is deeper, $m_R
\sim 26.2$, which corresponds to smaller objects ($\sim 2.5$ smaller
diameter) and will be complete before the LSST begins full survey
operations.

\section{Guiding Principles of the Survey} 
\label{principles}

To constrain our overall survey design, we adopted principles derived
from our science goals. The main science goals are discussed in more
detail by \cite{2019EPSC...13..395T} but in summary are: (1) measure
the size distribution of the TNOs down to 25 km, (2) measure the
colors of about one thousand TNOs, (3) derive the shape distribution
of TNOs from partial lightcurves, and (4) measure colors, size
distribution, and shapes as a function of dynamical class and size.

Additional secondary science goals are related to other
populations such as the main belt asteroids, Centaurs and Trojan
asteroids.

The guiding principles derived from these science goals follow:

\begin{itemize}

\item Survey depth is more important than measuring color. That is
  because the survey’s main goal is to probe small sizes and discover
  large numbers of objects.

\item Uniform depth per observing run per field is the most important
  feature of the survey. This will allow the ability to track objects
  consistently throughout the survey.

\item Colors will only be measured after uniform depth is reached in
  the $V\!R$ filter. This is because of the prior two principles and
  also because if objects' orbits are well-known, colors can be
  obtained in later epochs if need be.

\item We want to observe each object twice in the first year, once
  near opposition and once a month from opposition. Each object will
  then be observed in years 2 and 3 as well. The inclusion of the
  off-opposition observation allows for a better orbit solution than
  purely opposition observations, with a factor 10 lower astrometric
  error after the survey is completed (see Figure~\ref{orbits}).

\item We will track objects over the multi-year course of the survey,
  so we must expand our sky coverage as objects with different orbits
  disperse from their initial Year 1 discovery locations
  (Figure~\ref{decamplot}).

\item Objects in the outer solar system, beyond 30 au, are the primary
  goal, but science cases involving closer objects, including the main
  belt asteroids, will also be probed.

\end{itemize}

\begin{figure}[htbp]
\plotone{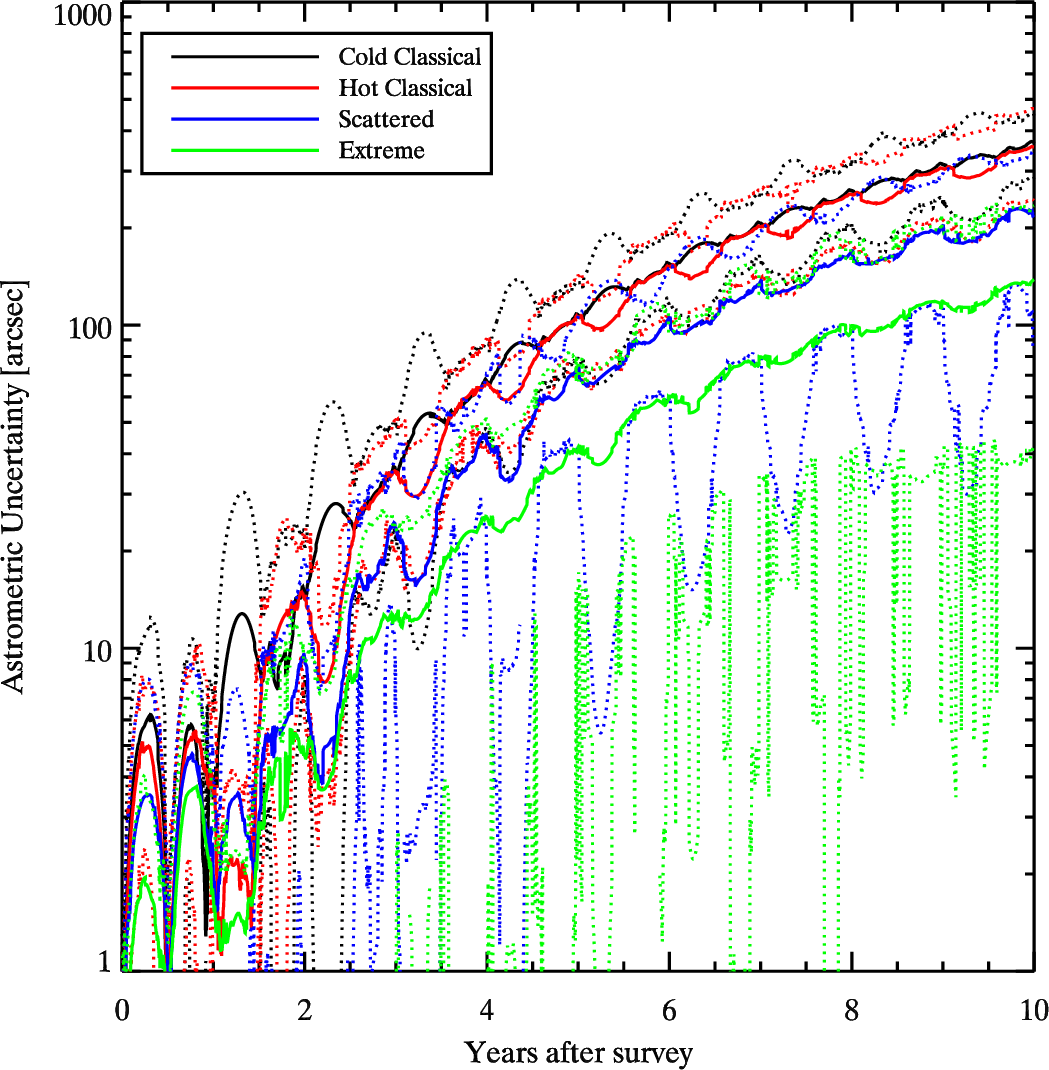}
\caption{Astrometric uncertainty after our survey is completed for
  different classes of TNOs. Simulated orbits were used to create
  simulated observations with the addition of Gaussian noise of 0.6
  arc-seconds in sky position similar to or greater than our actual
  astrometric uncertainty. Orbits were then fit to the simulated
  observations using the technique of
  \cite{2000AJ....120.3323B}. Object location predictions after survey
  completion were then compared to true simulated observations to
  determine astrometric uncertainty. Solid lines mark the median
  astrometric error while dashed lines bound the 90\% confidence
  region for each orbit type. Most populations maintain $< 30$
  arc-second error a few years after survey completion. \label{orbits}}
\end{figure}

\begin{figure}[htbp]
\plotone{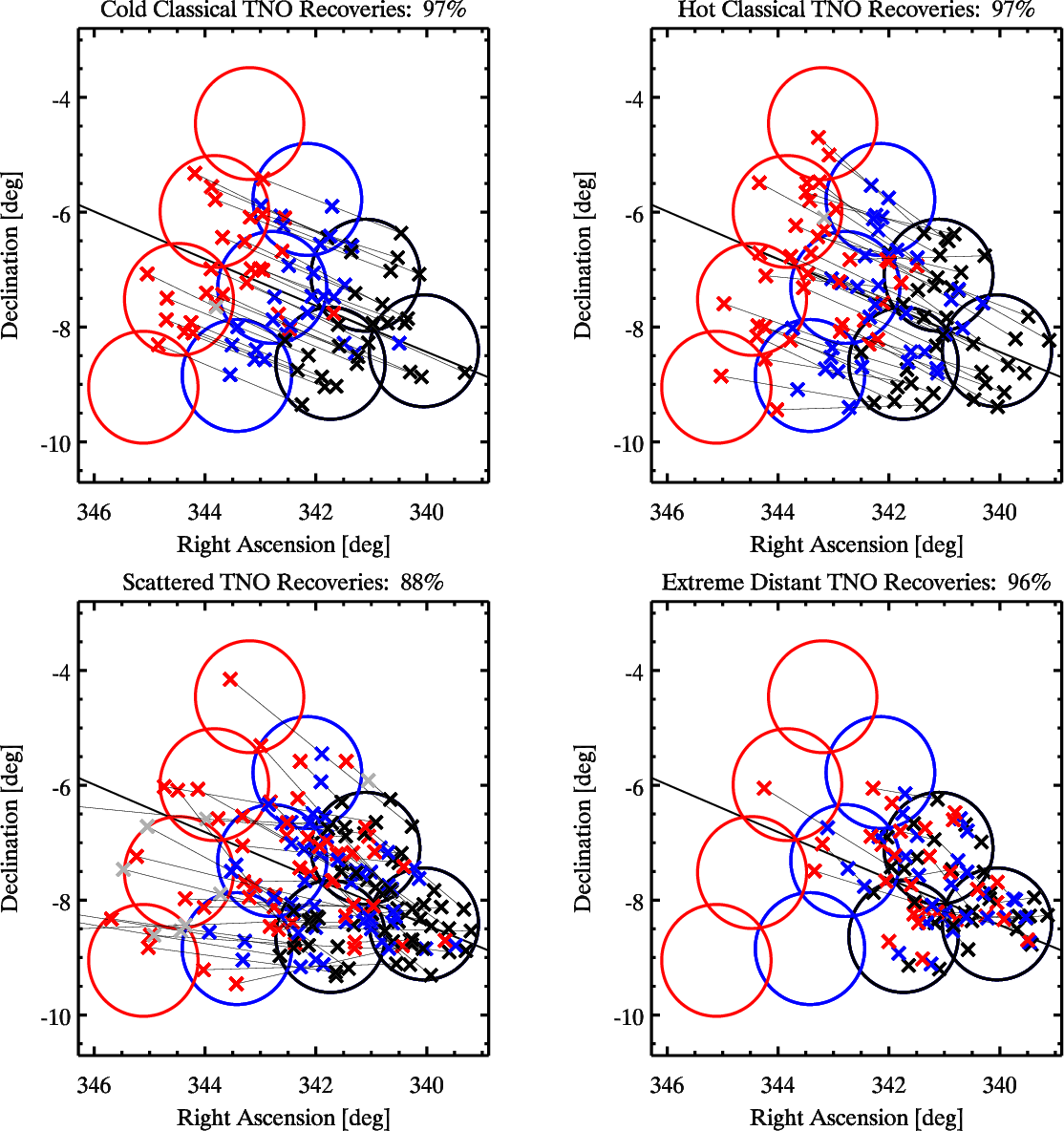}
\caption{A visualization of our recovery efficiency is shown. Circles
  approximate the DECam field positions. Black fields (rightmost 3
  fields) were imaged in Years 1–3, blue (center 3) in Years 2–3 and
  red (leftmost fields) in Year 3. Four populations of TNOs were
  modeled with simulated discovered objects shown as crosses with
  color coding for the year they are imaged. Grey crosses represent
  objects that are not imaged in a given year and grey lines connect
  object positions across epochs. Recovery efficiency is nearly 90\%
  or more for all classes of objects. The numbers of plotted objects
  have been reduced for clarity to depict orbital motion. In reality
  many thousands of objects were simulated and are expected to be
  discovered in the survey. The black line represents the solar system
  invariable plane as defined by \cite{2012jsrs.conf..231S}.}
\label{decamplot}
\end{figure}

\section{Observing Strategy}
\label{observing}

Following the guiding principles, our observing strategy is summarized
here, including the selection of the instrumentation for the
experiment, the exposure times, filter choices and field locations.

\subsection{Instrumentation}

The basic choice we made for the instrument was the prime focus DECam
camera on the Blanco 4m telescope at Cerro Tololo. The telescope
itself is historic, completed in 1976 as a southern counterpart to the
Kitt Peak Mayall 4m telescope. With the recent upgrade of
instrumentation at the Blanco 4m telescope to the Dark Energy Camera
(DECam), the DECam/Blanco instrument combination became one of the
leading survey instruments in the world \citep{2018ApJS..239...18A}.

We were granted a large survey program using DECam to survey our
region over a three year period. Given the amount of telescope time
devoted to the survey, $\sim 50$ observation dates, most of which were
half-nights, we expect that the findings of the survey will remain
relevant for a decade or longer.

\subsection{Exposure Times}

Exposure times must be relatively short so that asteroids do not trail
much given their apparent rate of motion and the typical image
quality, one of our secondary science goals. Typical asteroid motion
at heliocentric distances from 2 au to 3.5 au at opposition is 43 to
28 arc-seconds/hour \citep{1996AJ....112.1225J}. In the best seeing
conditions for the DECam instrument, 0.8 arc-seconds, the fastest
asteroids will not trail with exposure times of 0.8 arc-seconds / (43
arc-seconds / hour) x 3600 seconds / hour = 67 seconds. Readout time
for DECam is about 20 s, so choosing such an exposure time would
equate to an open shutter efficiency of 77\% (0.14 magnitudes loss
compared to 100\%). The optimum exposure time is therefore a trade-off
between limiting magnitude for the asteroids, which suggests shorter
exposures, compared to open-shutter efficiency which affects the
limiting magnitude for the slower objects, which suggests longer
exposures.

To strike a balance between those two competing factors, we have
chosen an exposure time of 120 seconds. This yields efficiency due to
the readout time of 86\% (0.08 magnitude loss for slow objects). In
terms of trailing loss (applicable to the fastest objects) this would
be 0.17 to 0 magnitudes trailing loss for slowest asteroids (28
arc-seconds per hour) in seeing of 0.8 arc-seconds to 1.1
arc-seconds. And this exposure time yields 0.6 to 0.3 magnitudes
trailing loss for fastest asteroids (43 arc-seconds per hour) in
similar seeing conditions.

\subsection{Expected Survey Depth in $V\!R$ and Predicted Discovery Numbers}
\label{expected}

We can estimate the survey depth based on the performance achieved
during the Dark Energy Survey (DES) DR1
\citep{2018ApJS..239...18A}. The bandpass of our chosen filter, the
$V\!R$ filter, is not specifically studied in
\cite{2018ApJS..239...18A}, but is roughly equivalent in terms of
total wavelength breadth and somewhat superior in overall throughput
to that of $g$ and $r$ combined from the DES DR1 (see
Section~\ref{colors}). The overall sensitivity of a single 90 second
exposure in $g$ and $r$ was estimated to be $g = 23.57$ and $r =
23.34$ for signal-to-noise ratios of 10 ($S/N=10$) for the DES
DR1. Since $S/N$ is proportional to the square root of exposure time
$t$, and similarly the square root of overall bandpass $b_p$, as $S/N
\sim \sqrt{b_p}$, we expect our limiting magnitude $m_{R}$ from filter
change alone to be roughly $m_r \approx 0.5 \times (23.34 + 23.57) +
2.5 \log(\sqrt{2}) = 23.83$.

In addition, we expect to take about $\sim 100$ images each of 120
seconds compared to the 90 second single images discussed in
\cite{2018ApJS..239...18A}. Also, we only need $S/N = 6.5$ for
detection, as compared to the DES DR1 $S/N = 10$. We expect that our
combination of images will not be as efficient as the static sky DES
found primarily because we are canvassing many possible apparent rates
of objects in our analysis. Not only this, but since we plan to
subtract a static image template, in some cases residual flux remains
from slowly moving objects. The theoretical magnitude increase for the
full DES survey, which targets about 10 images per sky location, would
be $1.25 \log(10) = 1.25$ magnitudes. The actual realized performance
appeared to be about 0.5 magnitudes less than this as the mean
survey-wide magnitude limits for $g$ and $r$ are 24.33 and 24.08
respectively (for the co-add magnitude limit with 1.95 arc-seconds
diameter, $S/N=10$), which are 0.76 and 0.74 magnitudes better than
the single image depth, a loss of about 0.5 magnitude over the
theoretical maximum. We expect our multiple-image depth to be worse
than this since our objects are moving and are found by combining
images assuming apparent rates for possible objects since we do not
{\it a priori} know their apparent motion rates. Assuming that our
image combination loss to be about 0.75 magnitudes versus the DES
performance, we expect our limiting magnitude $m_{r}$ (which is
roughly equivalent to $m_{R}$) to follow:

\begin{eqnarray} \nonumber
m_{R} & \approx & 23.83 + 1.25 \log \frac{120 \mbox{ s}}{90 \mbox{ s}}
+ 1.25 \log \frac{100 \mbox{ exposures}}{1 \mbox{ exposure}} + 2.5
\log \frac{10}{6.5} -
0.75 \mbox{ image combination correction} \\ \nonumber
m_{R} & \approx & 26.21
\end{eqnarray}

This image depth of approximately $\sim 26.2$ is well beyond that
expected for the LSST single image depth, as well as any other prior
wide-field survey sensitive to TNOs.

Given the image depth of $\sim 26.2$, we can estimate how many objects
we might we discover at that depth. To date, no surveys have observed
this faint for this amount of sky area so this number is highly
uncertain. For this estimate, we choose a comparison survey that is of
similar depth since surveys that are not as sensitive suffer from
extrapolation and surveys that are fainter tend to have lower number
statistics. Perhaps the most relevant similar survey is that of
\cite{2009ApJ...696...91F} who used data from the Subaru 8.2 m
telescope to detect 82 TNOs to a depth of $R = 25.7$ with a sky area
of 3.57 square degrees near the ecliptic in a re-analysis of data
collected for a Uranian satellite search
\citep{2005AJ....129..518S}. These results yield an ecliptic density
of 23 objects per square degree. Since our survey is about 0.5
magnitude fainter, using the brightness distribution from
\cite{2009ApJ...696...91F} with the slope of the faint object apparent
magnitude distribution $\alpha = 0.3$ derived from all surveys at the
time, this magnitude difference would result in a correction factor of
$10^{0.3 \times 0.5} = 1.41$ times more objects, yielding an expected
ecliptic detection rate of 32 objects per square degree.

Using this number, we can extrapolate to our survey area. Here, we
make the distinction between full-survey objects (that is, objects
tracked throughout the entire survey and which have reliable orbital
information) and partial-survey objects (which, in the final planned
year would have distance and brightness information but uncertain
orbits). For the former population, we imaged 4 patches of sky
throughout the year for each of 3 fields. For the latter population,
we imaged 4 patches of sky throughout the year for each of 10
fields. Since the DECam camera is about 2.7 square degrees, this is an
effective sky area of between 32 and 108 square degrees. Thus, we
expect to discover about 1,000 objects with well-determined orbits and
3,500 objects with enough information to measure apparent magnitude
and radial distributions.

\subsection{Optimal Filters for TNO Colors}
\label{colors}

In order to assess the number of objects for which we could obtain
color information, we again compared our survey to the results of the
DES DR1 \citep{2018ApJS..239...18A}. From DR1, we could determine that
$g$ goes deeper than $r$ by 0.23 magnitudes and $g$ probes deeper than
$i$ by 0.79 magnitudes. Here we consider $g-V\!R$, $r-V\!R$ and
$V\!R-i$ as well as the more traditional $g-r$. From the DES DR1
filter performance and color information of TNOs, we can assess the
optimal filter system to use in terms of number of objects for which
color information can be obtained. Here, we report our observational
choices primarily for future works. Although we planned to estimate
TNO colors using DECam, weather loss and cancellation of observing
runs due to COVID-19 resulted in fewer observational nights realized
compared to our plans. Therefore, in actuality, although we planned
our survey to measure TNO colors from DECam itself, we did not
implement this plan as our primary goal was to achieve very deep
$V\!R$ depth.

\subsubsection{Colors with $g-V\!R$ versus $g-r$}

The first question is whether $g-V\!R$ is superior to $g-r$. The
$V\!R$ filter is not examined in the DES DR1 paper because its total
throughput with the telescope is not measured with the analysis tools
used for DES DR1. However, we believe the $V\!R$ bandpass can be
estimated from the DR1 paper sensitivity. A summary of filter bandpass
in terms of approximate mean wavelengths and throughputs is described
in the DES DR1 paper and the $V\!R$ filter bandpass information is
found on the DECam
webpages \footnote{\url{https://noirlab.edu/science/programs/ctio/filters/Dark-Energy-Camera/VR-filter}}:

\begin{itemize}

\item $g$ central wavelength = 482.0 nm (416.5 nm - 547.5 nm, average
  peak 50\% throughput)

\item $r$ central wavelength = 641.25 nm (567.0 nm - 715.5 nm, average
  peak 80\% throughput)

\item $V\!R$ central wavelength = 626.5 nm (496.5 nm - 756.5 nm,
  estimated average peak 80\% throughput)

\item $i$ central wavelength = 782 nm (708.0 nm - 856 nm, average peak
  95\% throughput)

\end{itemize}

The throughput and filter bandpass information gives us an approximate
measure of the total number of photons falling on the
detector. Multiplying bandpass (nm) by throughput (a unitless
fraction) yields $g$ = 65 nm, $r$ = 118 nm, $V\!R$ = 208 nm, and $i$ =
140 nm. So achieving the same signal-to-noise in $r$ compared to
$V\!R$ would require about 208 nm / 118 nm = 1.76 times the exposure
time (assuming similar sky background which seems reasonable since
they have the same central wavelength). This overhead would result in
a magnitude loss of about $2.5 \log \sqrt{1.76}$ = 0.31
magnitudes. The faint TNO size distribution of
\cite{2004AJ....128.1364B} is roughly $10^{\alpha \Delta m}$ where
$\alpha = 0.32$ and $\Delta m$ is the magnitude limit between two sets
of observations, so this 0.31 magnitude loss corresponds to 25\% fewer
objects. It might be tempting to still choose $r$ over $V\!R$ to
improve color measurements, but this is not necessary as this effect
is marginal since the central wavelength difference between $g-V\!R$
(147.5 nm) and $g-r$ (170 nm) is very close. In fact, $g$ measurements
are nearly decoupled from $V\!R$ as they only overlap by about 50 nm,
less than one third of the $V\!R$ bandpass. Thus, $g-V\!R$ is
comparable to $g-r$ in terms of object color sensitivity due to
central wavelength differences, but takes less time to execute at the
telescope given the wide bandpass of $V\!R$ compared to $r$.

\subsubsection{Colors with $V\!R-i$ versus $g-V\!R$}

Given that $g-V\!R$ appears operationally superior to $g-r$, we then
explored whether the TNOs are red enough to chose $V\!R - i$ rather
than $g-V\!R$ for color measurements.

The blue TNOs have $R-I \sim 0.45$ and $V-R \sim 0.47$ while the red
TNOs have $R-I \sim 0.65$ and $V-R \sim 0.65$, so $V-I \sim 0.9$ and
$1.3$ magnitudes for blue and red TNOs respectively \citep[Figure
  2,][]{2008ssbn.book...91D}. The color measurement of $V-I$ spans
about 332 nm between the central wavelengths of the two filters. The
difference between $g$ and $i$ central wavelengths is about the same
as the difference in $V-I$. The sky is also brighter in $i$ since the
DR1 paper shows that with similar exposure time, the depth in $i$ is
0.8 magnitudes worse than $g$. Therefore, for blue objects it does not
matter if we use $g$ or $i$. For the red objects, we are 0.5
magnitudes better off in $i$, which yields $10^{\alpha \Delta m}$
times more objects, where $\alpha = 0.32$, so 45\% more objects
\citep{2004AJ....128.1364B}. The overlap in wavelength of $i$ with
$V\!R$ is similar to $g$ and $V\!R$, so there is no exacerbating issue
that the two color measurements might have different sensitivities due
to bandpass overlap.

Although $i$ seems favored over $g$ for colors when paired with the
$V\!R$ filter, there are more difficulties with calibration in $i$ due
to fringing. However, the DES analysis pipeline has been extensively
tested with mitigating fringing so we expect that the public pipeline
could be used for this calibration of the data if needed. The use of
$i$ has additional advantages over $g$ --- it is less sensitive to the
moon and seeing is generally better in $i$.

Given the above factors, for this project we planned to measure colors
using $V\!R$ - $i$. The choice of $V\!R$ over $r$ was made because of
bandpass. The choice of $i$ over $g$ was made due to the redness of
the TNOs, which favors $i$ wavelengths. Although unconventional, the
$V\!R$ - $i$ color measurement should take about half the exposure
time compared to a more traditional $r$ - $i$ measurements because of
the much improved throughput of the $V\!R$ filter. This assumes that
color measurements would take place during survey time epochs, which
are primarily performed with the $V\!R$ filter. If color measurements
were to take place in a separate epoch, it could be efficient to use
$g-r$ or $r-i$ in a single epoch to minimize issues related with
lightcurves, which might preclude the linking of prior epoch $V\!R$
colors with future epoch color measurements.

\subsection{Field Layout and Orbit Determination}
\label{orbitdetermination}

One of the primary science goals of this study is to determine the
orbits of discovered objects by following their sky location over the
course of a few years. In observational terms, the key issue for our
survey is correctly linking the position of a particular object
throughout the observations without confusion from other objects that
happen to be nearby as well. Each DECam field is about 2 degrees in
diameter, and not quite circular, for a sky area of about 2.7 square
degrees. We expect in the first year to cover 3 of these fields,
extending to 10 fields by the final years of this survey. Our overall
discovery rate is expected to be at best a few thousand objects and
our sky area to be covered is about 8.1 square degrees per 3 field
patch in the first year, of which there are four patches distributed
about the ecliptic. To design the survey to perform for the most
difficult case, which is a larger than expected number of objects, we
adopted a working value of 3,000 objects discovered in the first year,
which is a more difficult case than the expected value of $\sim 1,000$
objects in Year 1 (Section~\ref{expected}). This suggests a maximum
object sky density of 8.1 square degrees x $60\arcmin/\arcdeg \times
60\arcmin/\arcdeg \times 4$ patches / 3,000 objects $\sim$ 39 square
arc-minutes per object. Although the base number, $\sim 3,000$
objects, may not have been realized, this was a best-case scenario
upon which we based our recovery plan on and is what follows here.
Although there is some overlap among the Year 1 fields, it is less
than 10\% of each field and varied from patch to patch due to the
ecliptic angle with respect to celestial north so we did not consider
the effect of the overlap at the planning stage.

Data processing is expected to be done at the epoch level --- that is,
objects will be identified in each field pointing each year using a
shift-and-stack methodology to identify objects much fainter than the
single image depth threshold
\citep{2019arXiv190108549J,2022AAS...24022705G}. These discoveries
will then be linked to objects found in subsequent years of
observation, to be identified in the same manner. One of the primary
issues for our survey is that of confusion -- the sky density of newly
discovered objects is high enough that they can be confused with other
objects in year-to-year observations. The sky density of objects is
about 1 per 39 square arc-minutes (see above). This suggests that
object locations must be predicted to an accuracy of about $\sqrt{39
  \rm{sq arcmin}} \approx 6\arcmin$ to avoid major confusion
issues. Another method to reduce confusion might be to split the $\sim
4$ hours of observations we collected each night across two nights. We
did not specifically explore or execute this possibility because it
would make the survey more sensitive to night-to-night weather
differences and would increase the computational burden of stacking
this longer baseline of images to find objects over a 24 hour timebase
rather than the 4 hour timebase we executed.

The exact accuracy of object predictions from year-to-year is a
function of many unknowns including object orbital properties (which
are not known as this is one of the goals of the survey), specific
observing windows allocated to the project on a semester-by-semester
basis, and survey depth achieved based on actual local conditions at
the time of observation. Simulating these largely unpredictable
effects on the survey is very difficult because of the large number of
free parameters. Instead, to determine an adequate observation
cadence, we studied the more simplistic question of whether object
locations could be identified in years subsequent to the original 3
year survey duration and to what accuracy, which implicitly assumes
that the objects can be traced through all epochs of observations
relevant to each object.

\subsubsection{Field Layout}

To optimize potential survey effectiveness, we simulated the apparent
motion of 4 populations of objects over the original 3 year course of
the survey: Cold (low inclination) Classical Trans-Neptunian Objects,
Hot (higher inclination) Classical TNOs, Scattered TNOs, and high
perihelion TNOs such as Sedna. The optimal survey pattern for
following these objects over the course of a few years appeared to be
increasing the sky area each year to follow the fastest objects found
in Year 1 while also retaining slower objects which never move beyond
the original three field Year 1 pattern. The circles in
Figure~\ref{decamplot} show our survey plan with approximate DECam
field positions. Black fields were imaged in Years 1–3, blue in Years
2–3 and red Year 3. Our survey was extended into Year 4, partly due to
the COVID-19 pandemic, which resulted in the decision to include an
additional five fields at the wide end of the ``fan'' pattern that
were not originally studied in our simulations.

We show four of the populations of TNOs with simulated discovered
objects in Figure~\ref{orbits} with recovery efficiency shown in each
title and is nearly 90\% or more for all classes of objects. The
numbers of detected objects in the plots are representative only in
order to depict orbital motion --- thousands of TNOs are expected to
be discovered in the survey and many more were simulated to assess
survey strategy.

\subsubsection{Orbit Determination}

We simulated typical expected orbital parameters for Cold (low
inclination) Classical Trans-Neptunian Objects, Hot (higher
inclination) Classical TNOs, Scattered TNOs, and high perihelion
TNOs. We then computed the positions of each of these objects through
a field observation pattern that included both opposition and
quadrature observations in the first year and only opposition
observations in the second and third years. Noise was added at typical
centroid uncertainties of 0.6 arc-seconds for each of the positions
computed. Then, these noise-added simulated observations were used to
fit an orbit using the orbit determination method of
\cite{2000AJ....120.3323B}. This new orbit based on noise-added
simulated observations was then predicted for future years (0 -- 10
years after survey completion) and compared to the true simulated
orbit. The difference for the prediction from orbits of the
noise-added simulated observations compared to the true simulated
orbit predictions was then used to estimate the accuracy for the
populations of interest. Full orbital parameters simulated are given
in Table~\ref{obssimparameters}.

Overall, we found that using the optimal cadence of one observation
near quadrature and one near opposition the first year, then only
opposition observations in years 2 and 3 resulted in median sky-plane
position accuracies of 10 arc-seconds or better for most populations
about 2 years after the conclusion of the survey and better than 90
arc-second astrometric uncertainty about 4 years after the survey
conclusion. The addition of the single epoch of quadrature
observations the first year was critical --- this inclusion lowered
astrometric uncertainty by about a factor of 10. This is expected as
for opposition observations both parallactic motion and intrinsic
orbital motion contribute to apparent sky-plane velocity and
quadrature observations break this degeneracy since parallactic motion
is minimized \citep{2000AJ....120.3323B}. This is well within the
nominal accuracy required to combat confusion described in
Section~\ref{orbitdetermination} of 6 arcminutes, so we expect that
confusion will not be an issue for the DEEP survey.

\begin{deluxetable}{llll}
\tablewidth{7.5 in}
\label{obssimparameters}
\tablecaption{Population parameters used for orbital
  simulations. These were used to characterize the behavior of
  known sub-populations of the TNOs to design the survey. Inclination
  parameters $\sigma$ and $\mu$ are described in
  \cite{2010AJ....140..350G}. The primary parameters of importance in
  survey design are the semi-major axis distributions of the
  populations, which constrain year-to-year motion of the objects, and
  the inclination distributions, which constrain the width of the
  ``fan''-shaped sky area covered.}
\tablehead{
\colhead{Parameter} & \colhead{Value} & \colhead{Population} & \colhead{Description} }
\startdata
\hline
& & Orbital Parameters for Orbital Simulations \\
\hline
$a_{\rm min}$ & 42 au  & Cold Classicals & Minimum semi-major axis \\
$a_{\rm max}$ & 45 au  & Cold Classicals & Maximum semi-major axis \\
$q_{\rm min}$ & 42 au  & Cold Classicals & Minimum perihelion \\
$q_{\rm max}$ & 45 au  & Cold Classicals & Maximum perihelion \\
$e_{\rm min}$ & 0.0  & Cold Classicals & Minimum eccentricity \\
$i_{\sigma}$ & 2.0\arcdeg  & Cold Classicals & Inclination width $\sigma$ \\
$i_{\mu}$ & 0.0\arcdeg  & Cold Classicals & Inclination mean $\mu$ \\
\hline
$a_{\rm min}$ & 42 au  & Hot Classicals & Minimum semi-major axis \\
$a_{\rm max}$ & 45 au  & Hot Classicals & Maximum semi-major axis \\
$q_{\rm min}$ & 42 au  & Hot Classicals & Minimum perihelion \\
$q_{\rm max}$ & 45 au  & Hot Classicals & Maximum perihelion \\
$e_{\rm min}$ & 0.0  & Hot Classicals & Minimum eccentricity \\
$i_{\sigma}$ & 8.1\arcdeg  & Hot Classicals & Inclination width $\sigma$ \\
$i_{\mu}$ & 0.0\arcdeg  & Hot Classicals & Inclination mean $\mu$ \\
\hline
$a_{\rm min}$ & 250 au  & Extreme TNOs & Minimum semi-major axis \\
$a_{\rm max}$ & 1250 au  & Extreme TNOs & Maximum semi-major axis \\
$q_{\rm min}$ & 50 au  & Extreme TNOs & Minimum perihelion \\
$q_{\rm max}$ & 500 au  & Extreme TNOs & Maximum perihelion \\
$e_{\rm min}$ & 0.65  & Extreme TNOs & Minimum eccentricity \\
$i_{\sigma}$ & 6.9\arcdeg  & Extreme TNOs & Inclination width $\sigma$ \\
$i_{\mu}$ & 19.1\arcdeg  & Extreme TNOs & Inclination mean $\mu$ \\
\hline
$a_{\rm min}$ & 50 au  & Scattered TNOs & Minimum semi-major axis \\
$a_{\rm max}$ & 150 au  & Scattered TNOs & Maximum semi-major axis \\
$q_{\rm min}$ & 30 au  & Scattered TNOs & Minimum perihelion \\
$q_{\rm max}$ & 40 au  & Scattered TNOs & Maximum perihelion \\
$e_{\rm min}$ & 0.65  & Scattered TNOs & Minimum eccentricity \\
$i_{\sigma}$ & 6.9\arcdeg  & Scattered TNOs & Inclination width $\sigma$ \\
$i_{\mu}$ & 19.1\arcdeg  & Scattered TNOs & Inclination mean $\mu$ \\
\enddata
\end{deluxetable}

Orbits will be well-determined by our plan. Using our simulation, we
also computed expected fractional accuracy for orbital elements from
our observing cadence and duration using noise-added astrometric
positions as described above, and are presented in
Table~\ref{orbits}. These were computed by measuring the absolute
magnitude of the difference between the true orbital quantity and the
estimated value determined by the method of
\cite{2000AJ....120.3323B}. This difference was then divided by the
true quantity to yield fractional error. For instance, for semi-major
axis with true and estimated values $a_{\rm true}$ and $a_{\rm est}$,
the fractional accuracy was $|a_{\rm true} - a_{\rm est}| / a_{\rm
  true}$, which in our simulations was 0.4\% (Table~\ref{orbits}).

We note that these are predicted orbital accuracies as actual orbital
accuracies can only be done on and object-by-object basis by running
dynamical simulations of many clones for millions of years --- a
process that is not tractable for a large suite of simulated objects
as we study here. Such an analysis will be done at a later time when
full knowledge of our discovered objects is known. Our predicted
accuracies are similar to, but a factor of a few worse than what was
reported by \cite{2018ApJS..236...18B} with a semi-major accuracy of
$\delta a / a \lesssim 0.1\%$ for the vast majority of their classical
non-resonant objects. This value was determined for their objects
after they were discovered and extensive simulations were run. We may
yet achieve that accuracy as our original survey was extended to a
timebase of 3 years (rather than our original 2 which is considered
here) due to the COVID-19 pandemic. In addition, the $1\sigma$
astrometric uncertainty given to observations was 0.6 arcseconds,
which was a deliberate overestimate of our estimated astrometric
uncertainty used for planning purposes.

\begin{deluxetable}{llll}
\tablewidth{7.5 in}
\label{orbits}
\tablecaption{Orbit accuracy based on our simulated object classes for
  our survey cadence and duration (observations in Year 1 at
  opposition and quadrature, then opposition in Years 2 and
  3). Fractional error is computed by calculating the absolute
  magnitude of the difference between the true quantity and the
  estimated orbital quantity and dividing by the true quantity, as
  described in the text. Orbital quantities are semi-major axis (in
  au), eccentricity and inclination (in degrees), denoted by $a$, $e$,
  and $i$.}  \tablehead{ \colhead{Class} & \colhead{quantity} &
  \colhead{median fractional error} & \colhead{median value} }
\startdata \hline Cold Classical & $a$ & 0.0039 & 43.40 \\ Cold
Classical & $e$ & 0.018\tablenotemark{a} & 0.012 \\ Cold Classical &
$i$ & 0.0044\tablenotemark{a} & 1.66 \\ Hot Classical & $a$ & 0.0043 &
43.33 \\ Hot Classical & $e$ & 0.014\tablenotemark{a} & 0.011 \\ Hot
Classical & $i$ & 0.0027 & 5.78 \\ Scattered & $a$ & 0.059 & 81.61
\\ Scattered & $e$ & 0.068 & 0.569 \\ Scattered & $i$ & 0.0037 & 19.60
\\ ETNO & $a$ & 0.43 & 479.16 \\ ETNO & $e$ & 0.11 & 0.801 \\ ETNO &
$i$ & 0.0038 & 20.32 \\ \enddata \tablenotemark{a}{For these
  quantities, the true value nears 0 for the simulated population, so
  the fractional error is poorly defined and instead the absolute
  error is reported, which is the absolute magnitude of the difference
  between the true and estimated values.}
\end{deluxetable}

\subsection{Field Considerations for A Semester}
\label{asemester}

Our initial observing windows in 2019A, the first semester of the
project, were UT Dates April 2--4, (second halves of the nights), May
5--8 (full nights), and June 2--4 (first halves of the
nights). Although we also had observing dates scheduled for July 7--9
(second halves), these are late enough in the year that they would be
accessing the B Semester fields, which are described in
Section~\ref{bsemester}. According to Horizons ephemeris estimates for
the Sun position \citep{1996DPS....28.2504G}, the opposition point of
the A semester observing dates were as follows (in J2000 RA and Dec):
April 2 12:43, -4:38; May 5 14:46, -16:02; June 2 16:37, -22:05. From
the CTIO ephemeris page the sidereal times between twilight were as
follows: April 2 08:00–17:37, May 5 09:39–20:05, June 2 11:18–22:09.

Since our only full nights were scheduled in May, and most of our
objects of interest have fairly uniform distributions in ecliptic
longitude, it made sense to choose the A semester fields based on the
allocated May 2019A time. Ideally, these would be within $\pm 1$ hour
of opposition. This is close enough to opposition that the apparent
motion is quite high and then in the month preceding, the -1 hour from
opposition fields will be at quadrature and the +1 hour from
opposition fields will also be at quadrature. Thus, the two groups of
fields should ideally be around 13:46 and 15:46. Pairing both
quadrature and opposition observations in the same year results in
superior constraints on orbits as the opposition observations probe
heliocentric distance from parallactic motion which are combined with
orbital motion while the quadrature observations probe primarily
orbital motion \citep{2000AJ....120.3323B}, as discussed in
Section~\ref{orbitdetermination}.

Minor adjustments were made to the above based on the fact that the
ecliptic is not parallel to the celestial equator, so the first sets
of fields in a full night don't rise as high in the sky as the second
sets of fields. Also, the first 2 hours of the nights in May were
scheduled for a different program so we did not consider those hours
for execution of our program. The goal was to equalize the depth of
the fields during each of the two visits. To do this, we chose
positions to equalize the time each field was above 1.4 airmasses and
when the Sun was below -15 degrees (halfway between nautical and
astronomical twilight boundaries).

Most of the populations of interest are thought to be relatively
uniform in ecliptic longitude (classical TNOs, Scattered TNOs,
Non-Hilda Asteroids, Centaurs). Exceptions to this are the Resonant
TNOs, Trojan Asteroids and the Extreme TNOs which may be affected by
the action of a distant massive planet
\citep{2014Natur.507..471T}. The four most populated Neptune
resonances, in order of total number of observed objects in each
resonance, are the 3:2, 2:1, 5:2 and 4:3. Examples of orbits of
objects in the 3:2, 2:1 and 4:3 resonances can be seen in
\citep[Figures 3--10,][]{1996AJ....111..504M}.

The most important point of the longitudinal distributions of the
resonant objects is that the 3:2 objects, which are by far the most
numerous, librate with their perihelia about 90 degrees from Neptune,
with a spread of about $\pm 15$ degrees. The 2:1 objects can be found
at many longitudes without introducing significant bias and the 4:3
come to perihelion 60 degrees from Neptune. So ideal field locations
would probe the area about 90 degrees from Neptune. This would allow
enhanced sensitivity to the smallest TNOs which would be the
Neptune-crossing 3:2 objects near perihelion. A secondary field could
be done 60 degrees from Neptune which would allow the 4:3 objects to
be probed and provide some sensitivity to Neptune Trojans. The benefit
of catching the 3:2 objects at perihelion versus aphelion is, for an
object with eccentricity $e$, in terms of magnitude is $2.5 \log
(\frac{1+e}{1-e})^4 = 10 \log \frac{1+e}{1-e}$. So for a 3:2 TNO with
a typical eccentricity of 0.23, the gain is $\Delta m$ = 2.0
magnitudes in brightness. In terms of object numbers, for shallow size
distributions such as the \cite{2004AJ....128.1364B} faint
Hubble-detected objects, the total number of objects is proportional
to $10^{\alpha \Delta m}$ where $\alpha = 0.32$, so this is about a
factor 4 in terms of object numbers. This also means we can probe
objects that are about a factor 2.5 smaller in size using this
methodology. That said, eccentric bodies do spend more time at
aphelion than perihelion, so there will be an effect on survey
sensitivity due to this issue, which we have not considered here.

Since the 3:2 TNOs have perihelion $90 \pm 15$ degrees ($6 \pm 1$
hours) from Neptune and the 4:3 TNOs are $60 \pm 15$ and $180 \pm 15$
degrees ($4 \pm 1$ and $12 \pm 1$ hours) from Neptune, this would
suggest the optimum field locations should be about 05:15 and 17:15
for the 3:2 objects and 04:15, 11:15 and 19:15 for the 4:3
objects. For the 2019A semester, it was difficult to reach these
resonant locations at opposition, but we could image one of the 3:2
locations. The 2019B semester provided a better opportunity although,
in the end, we chose to optimize fields based on observing date and
opposition proximity in the B semester (Section~\ref{bsemester}). The
Extreme TNOs \citep{2019AJ....157..139S} appear to come to perihelion
in the 3 to 6 RA hour range, so they were unavailable until
mid-2019B. Additionally, our main science goals are related to the
general TNO population which has very large expected discovery
numbers, without an undue focus on the Extreme TNOs, which are much
more difficult to detect \citep{2014Natur.507..471T}, so we decided to
not alter our survey fields just for the few Extreme TNOs that we
might find in the survey fields.

In summary, the actual field locations were chosen to lie along the
invariable plane as defined by \cite{2012A&A...543A.133S} and designed
such that equal amounts of time could be spent on each field during
the on-opposition and off-opposition observations given our 2019 A
semester observation dates. A similar methodology was followed for the
2019 B field choices. Subsequent years after 2019 were solely based on
our initial 2019 field choices.

\subsubsection{Actual Field Coordinates Selected for 2019 A Semester}
\label{fan}

\begin{figure}[htbp]
\plotone{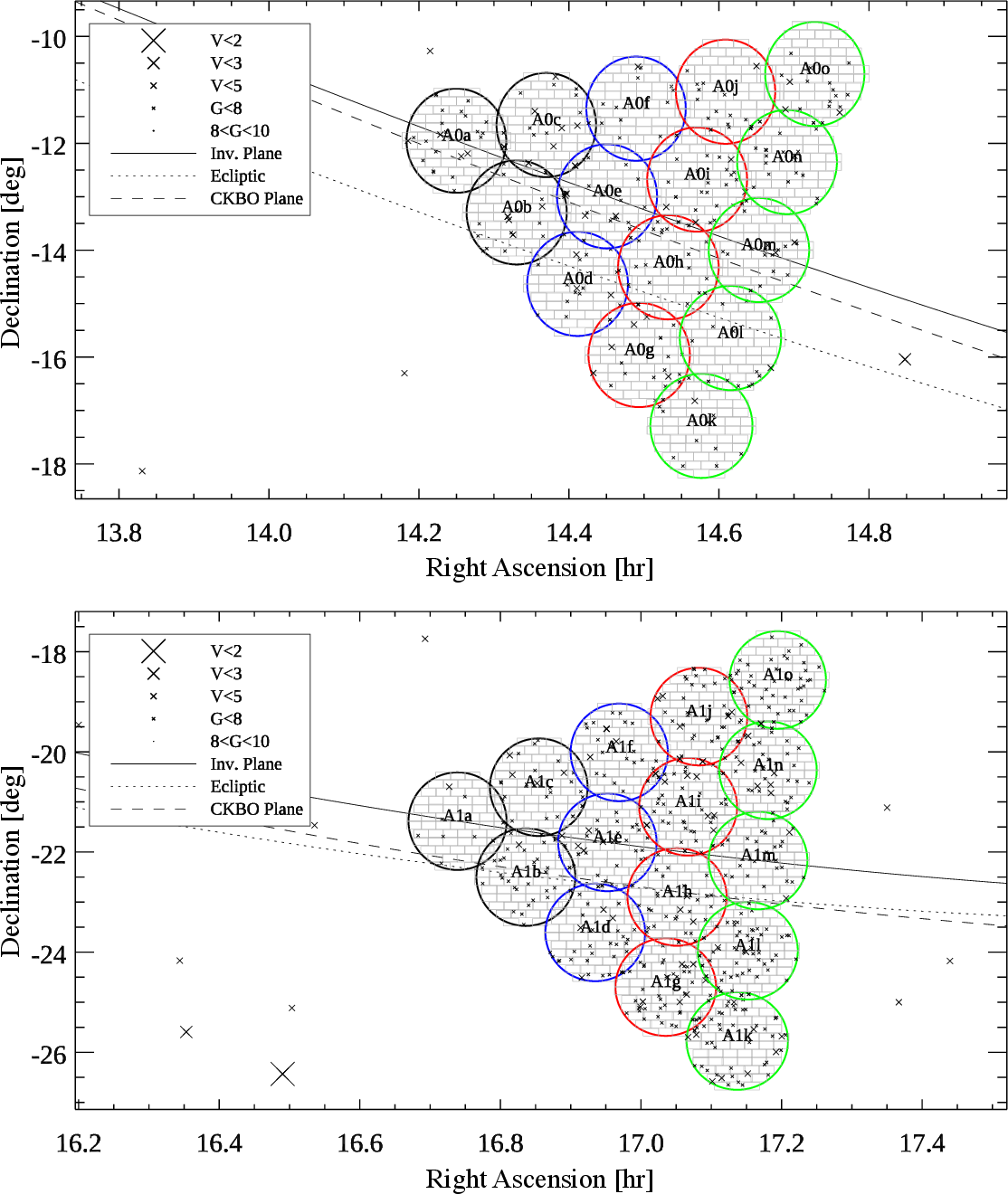}
\caption{The ``A'' semester fields, observed in the first half of each
  year of the survey. In year 1 the black fields (A0a-c and A1a-c)
  were observed. In year 2, the black fields were repeated and the
  blue fields (A0d-f and A1d-f) were added. In year 3, all year 2
  fields in addition to the red fields (A0g-j and A1g-j) were
  observed. The year 4 fields (A0k-o and A1k-o), in green, were not
  originally anticipated as part of the survey but were added later as
  mitigation for telescope time losses due to operational shutdowns
  related to the COVID-19 pandemic.}
\label{decamfieldsa}
\end{figure}

\begin{figure}[htbp]
\plotone{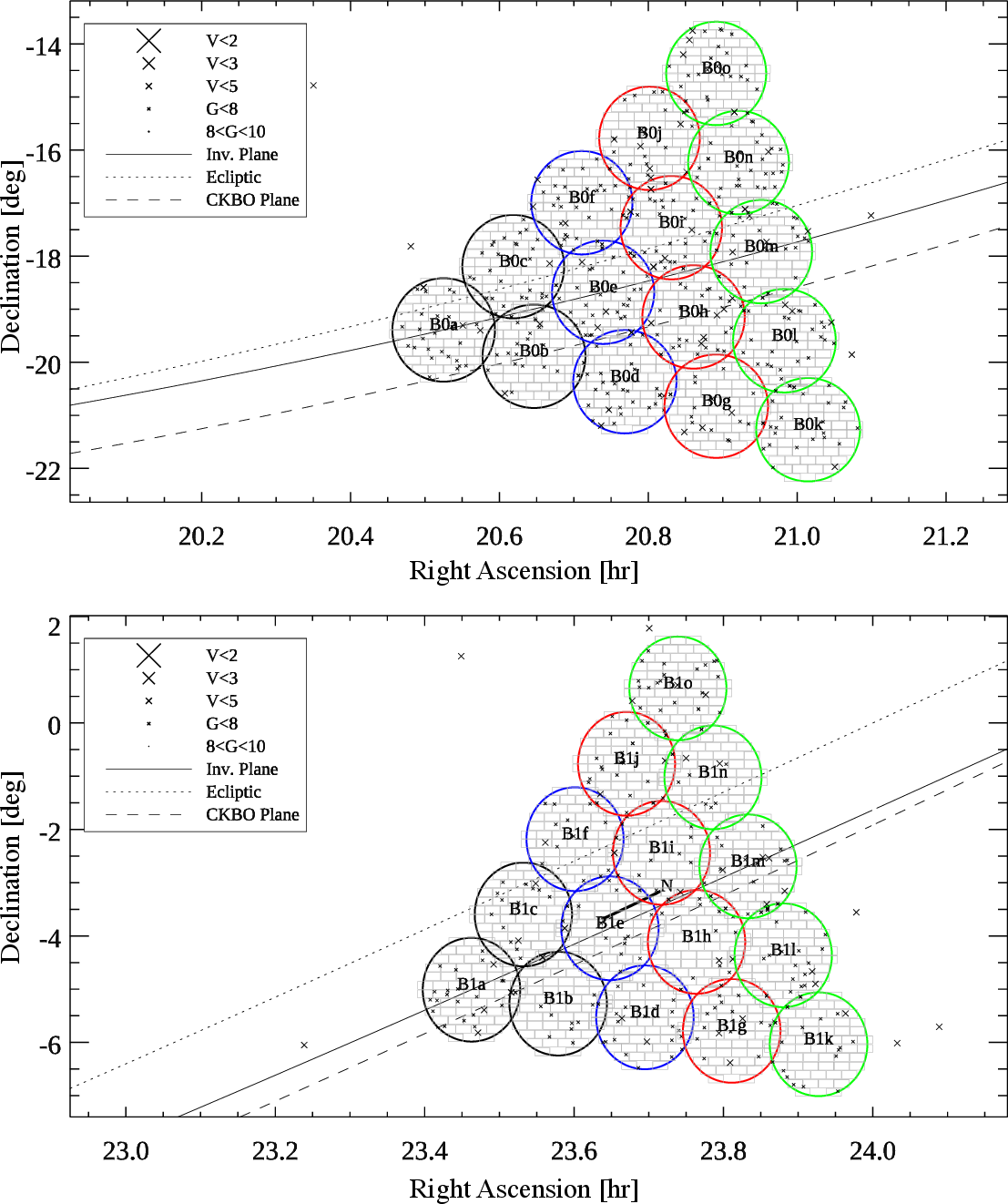}
\caption{The ``B'' semester fields, observed in the second half of
  each year of the survey. In year 1 the black fields (B0a-c and
  B1a-c) were observed. In year 2, the black fields were repeated and
  the blue fields (B0d-f and B1d-f) were added. In year 3, all year 2
  fields in addition to the red fields (B0g-j and B1g-j) were
  observed. The year 4 fields (B0k-o and B1k-o), in green, were not
  originally anticipated as part of the survey but were added later as
  mitigation for telescope time losses due to operational shutdowns
  related to the COVID-19 pandemic. Neptune's location though our
  2022B allocated observing window is marked with an N and a line.}
\label{decamfieldsb}
\end{figure}

Based on the approximate field centers above, we examined star
locations from the Gaia catalog to select fields minimizing the number
of stars brighter than $G<8$
\citep{2016A&A...595A...1G,2018A&A...616A...1G} and in particular the
especially bright stars with $V<5$. We have examined the final year 4
field configuration which is 15 full fields. The sky plots of the
fields are pictured in Figure~\ref{decamfieldsa} and
\ref{decamfieldsb}. All fields were moved less than 0.5 hours along
the invariable plane to avoid the brightest stars. We used the
invariable plane defined by \cite{2012jsrs.conf..231S} ($i=23.008861
\arcdeg$ and $\Omega = 3.8526111 \arcdeg$ with respect to the
celestial ICRF plane and $i = 1.5786944 \arcdeg$ and $\Omega =
107.58222 \arcdeg$ with respect to the J2000
ecliptic). \cite{2017AJ....154...62V} found that the classical TNOs
had $1 \sigma$ limits of $i=1.2$--$2.2\arcdeg$ and $\Omega = $63--$95
\arcdeg$, which is less than the maximum deviation from the invariable
plane of our field locations especially considering the cold classical
TNOs have an inclination distribution that is $\sin(i)$ times a
Gaussian with $1 \sigma = 2 \arcdeg$ \cite{2010AJ....140..350G}. More
recently, \cite{2023AJ....165..241M} provided improved estimates of
the Kuiper Belt mean plane, which for many heliocentric distances,
fall within our original design criteria. The actual field coordinates
are given here for our 2019 observing epoch. Note that at each
observation epoch in subsequent years, field coordinates were adjusted
based on the day of the year of observations and the expected apparent
motion of TNOs to minimize the loss of objects moving off of field
edges due to apparent motion caused by the Earth. Any orbital motion
of the TNOs was not accounted for in this manner because the overall
field ``fan shape'' was selected to adjust for the orbital motion
expected for the objects.

In 2019, we initially planned for 3 fields where all detected objects
would be further followed in our years-long survey. As the actual
fields observed each year were adjusted based on typical TNO apparent
motion, and the total duration of the survey was extended due to
COVID, the final field selection was not set until the 2022
observations. These field coordinates are shown for the 2022A semester
in Table~\ref{afields}.

\subsection{Actual Field Coordinates Selected for 2019 B Semester}
\label{bsemester}

The choice of B semester fields was largely based on our allocation of
telescope time in the first semester of observations following the
same methodology of the A semester observations. Given that we were
allocated telescope time in 2019 in late August and late September, we
planned our fields so that both the opposition and quadrature
observations were available during those time periods --- for the
early fields, opposition in late August and quadrature in late
September, and for the late fields, quadrature in late August and
opposition in late September. Given these constraints, plus the goal of
avoiding bright stars and the fact that the full field pattern of 15
fields wasn't determined until 2022, we report the 2022 epoch fields
in Table~\ref{bfields} for the B semester.

\begin{longrotatetable}
\label{observations}
\begin{deluxetable}{|l|l|l|l|l|l|l|l|l|l|l|l|}
  \tablecaption{Summary of field observations for the DEEP project as
    of 2022B. Sky coordinates are given as right ascension ($\alpha$)
    and declination ($\delta$). The value $t_{\rm eff} \approx 1$ is
    an empirical measure of efficiency roughly consistent with seeing
    full-width-at-half-maximum of 0.95 arc-seconds under clear,
    moonless skies -- optimal conditions with DECam, which is
    described in detail in Equation 4 of \cite{2018PASP..130g4501M}.
    $N$ is number of 120 s observations, $\chi$ is mean airmass,
    $\theta$ is mean seeing full-width-at-half-maximum, $m_{\rm sky}$
    is mean sky brightness, and $m_{\rm ext}$ is mean
    extinction. Computations were performed with algorithms supplied
    as part of the Dark Energy Survey Image Processing Pipeline
    \citep{2018PASP..130g4501M}. }

\tablehead{
\colhead{Local Date}     & \colhead{Night}    & \colhead{UT Time} & \colhead{Field} & \colhead{$\alpha$}  & \colhead{$\delta$}  & \colhead{$N$} & \colhead{$t_{\rm eff}$} & \colhead{$\chi$} & \colhead{$\theta$}  & \colhead{$m_{\rm sky}$} & \colhead{$m_{\rm ext}$} \\
\colhead{Start of Night} & \colhead{Fraction} & \colhead{Range}   & \colhead{Name}  & \colhead{[\arcdeg]} & \colhead{[\arcdeg]} & \colhead{}    & \colhead{}              & \colhead{}       & \colhead{[\arcsec]} & \colhead{}              & \colhead{} }
\startdata
2019/04/01 & second half  &     05:13--05:18,09:36--09:41 & A0c & 216.5 & -12.0 & 6      & 2.32      &    1.29 & 1.52 &  0.02 &        0.00 \\
2019/04/01 & second half  &     05:21--09:22              & A0a & 214.7 & -12.3 & 98     & 53.33     &    1.14 & 1.37 &  0.00 &        0.00 \\
2019/04/01 & second half  &     05:06--05:11,09:25--09:33 & A0b & 215.9 & -13.6 & 7      & 2.78      &    1.28 & 1.49 &  0.01 &        0.07 \\
2019/04/02 & second half  &     05:09--09:22              & A0b & 215.9 & -13.6 & 103    & 65.19     &    1.12 & 1.27 &  0.00 &        0.00 \\
2019/04/02 & second half  &     04:53--04:58,09:24--09:30 & A0a & 214.7 & -12.3 & 6      & 3.32      &    1.29 & 1.35 &  0.00 &        0.00 \\
2019/04/02 & second half  &     05:01--05:06,09:32--09:37 & A0c & 216.5 & -12.0 & 6      & 3.39      &    1.30 & 1.33 &  0.00 &        0.00 \\
2019/04/03 & second half  &     05:07--09:24              & A0c & 216.5 & -12.0 & 104    & 31.37     &    1.14 & 1.70 &  0.07 &        0.00 \\
2019/04/03 & second half  &     04:47,09:28--09:37        & A0a & 214.7 & -12.3 & 5      & 0.84      &    1.45 & 2.44 &  0.33 &        0.00 \\
2019/04/03 & second half  &     05:00--05:05,09:39--10:00 & A0b & 215.9 & -13.6 & 12     & 1.94      &    1.47 & 2.16 &  0.57 &        0.15 \\
2019/05/03 & full night   &     01:02--04:38              & A0a & 214.7 & -12.3 & 88     & 51.40     &    1.16 & 1.02 &  0.40 &        0.00 \\
2019/05/03 & full night   &     00:47--00:52,04:41--04:46 & A0b & 215.9 & -13.6 & 6      & 3.39      &    1.31 & 1.04 &  0.44 &        0.00 \\
2019/05/03 & full night   &     00:55--01:00,04:48--04:53 & A0c & 216.5 & -12.0 & 6      & 3.00      &    1.31 & 1.10 &  0.42 &        0.00 \\
2019/05/03 & full night   &     06:15--09:17              & A1a & 251.6 & -21.5 & 74     & 54.28     &    1.07 & 0.93 &  0.37 &        0.00 \\
2019/05/03 & full night   &     06:00,09:49--09:54        & A1b & 253.1 & -22.6 & 4      & 1.92      &    1.27 & 1.04 &  0.58 &        0.00 \\
2019/05/03 & full night   &     06:07--06:12,09:19--09:46 & A1c & 253.4 & -20.8 & 15     & 6.30      &    1.23 & 1.05 &  0.59 &        0.25 \\
2019/05/04 & full night   &     00:39--00:44,04:40--04:45 & A0a & 214.7 & -12.3 & 6      & 2.94      &    1.32 & 1.27 &  0.16 &        0.00 \\
2019/05/04 & full night   &     00:47--00:52,04:48--04:53 & A0b & 215.9 & -13.6 & 6      & 2.98      &    1.30 & 1.23 &  0.20 &        0.04 \\
2019/05/04 & full night   &     00:54--04:38              & A0c & 216.5 & -12.0 & 91     & 52.84     &    1.18 & 1.13 &  0.10 &        0.00 \\
2019/05/04 & full night   &     06:01--06:06,09:38--09:43 & A1a & 251.6 & -21.5 & 5      & 1.70      &    1.19 & 1.29 &  0.29 &        1.23 \\
2019/05/04 & full night   &     06:09--06:14,09:46--09:50 & A1b & 253.1 & -22.6 & 6      & 1.54      &    1.18 & 1.26 &  0.42 &        2.34 \\
2019/05/04 & full night   &     06:16--09:36              & A1c & 253.4 & -20.8 & 70     & 4.88      &    1.09 & 1.23 &  0.49 &        2.49 \\
2019/05/05 & full night   &     no data collected         & --- & ---   & ---   & ---    & ---       &    ---  & ---  &  ---  &        ---  \\
2019/05/06 & full night   &     00:52--00:57,04:40--04:45 & A0a & 214.7 & -12.3 & 6      & 5.29      &    1.26 & 1.00 &  0.09 &        0.00 \\
2019/05/06 & full night   &     01:07--04:38              & A0b & 215.9 & -13.6 & 86     & 23.68     &    1.13 & 0.99 &  0.22 &        1.22 \\
2019/05/06 & full night   &     00:59--01:04,04:48--04:53 & A0c & 216.5 & -12.0 & 6      & 5.82      &    1.26 & 0.97 &  0.07 &        0.00 \\
2019/05/06 & full night   &     05:58--06:03,09:40        & A1a & 251.6 & -21.5 & 4      & 0.11      &    1.11 & 1.10 &  0.40 &        3.29 \\
2019/05/06 & full night   &     06:13--09:37              & A1b & 253.1 & -22.6 & 79     & 27.93     &    1.09 & 1.03 &  0.37 &        1.09 \\
2019/05/06 & full night   &     06:06--06:11,09:47        & A1c & 253.4 & -20.8 & 4      & 0.07      &    1.12 & 1.02 &  0.42 &        3.13 \\
2019/06/01 & first half   &     00:09--00:14              & A1a & 251.0 & -21.4 & 3      & 0.69      &    2.12 & 1.59 &  0.47 &        0.00 \\
2019/06/01 & first half   &     00:24--04:38              & A1b & 252.5 & -22.5 & 99     & 63.78     &    1.29 & 1.09 &  0.21 &        0.00 \\
2019/06/01 & first half   &     00:16--00:21              & A1c & 252.8 & -20.7 & 3      & 0.78      &    2.13 & 1.47 &  0.49 &        0.00 \\
2019/06/02 & first half   &     23:25--23:33              & A0a & 213.6 & -11.9 & 4      & 1.40      &    1.35 & 0.92 &  1.10 &        0.00 \\
2019/06/02 & first half   &     00:40--04:37              & A1a & 251.0 & -21.4 & 96     & 71.78     &    1.22 & 1.04 &  0.11 &        0.00 \\
2019/06/03 & first half   &     23:31--23:58              & A0a & 213.6 & -11.9 & 12     & 7.31      &    1.28 & 0.98 &  0.35 &        0.00 \\
2019/06/03 & first half   &     00:04--00:09,04:22--04:29 & A1a & 251.0 & -21.4 & 8      & 2.97      &    1.41 & 1.43 &  0.17 &        0.00 \\
2019/06/03 & first half   &     00:14--00:16,04:34--04:39 & A1b & 252.5 & -22.5 & 6      & 2.40      &    1.52 & 1.31 &  0.26 &        0.05 \\
2019/06/03 & first half   &     00:19--04:19              & A1c & 252.8 & -20.7 & 97     & 28.39     &    1.31 & 1.46 &  0.21 &        0.41 \\
2019/07/06 & second half  &     04:49--07:17              & B0a & 308.8 & -19.2 & 60     & 2.11      &    1.04 & 1.37 &  0.23 &        2.33 \\
2019/07/06 & second half  &     04:33--04:39              & B0b & 310.6 & -19.7 & 3      & 1.29      &    1.13 & 1.52 &  0.00 &        0.13 \\
2019/07/06 & second half  &     04:41--04:46              & B0c & 310.2 & -18.0 & 3      & 0.58      &    1.11 & 1.78 &  0.01 &        0.44 \\
2019/07/07 & second half  &     05:11--08:32              & B0a & 308.8 & -19.2 & 80     & 2.65      &    1.06 & 1.44 &  0.09 &        2.15 \\
2019/07/07 & second half  &     04:55--05:01              & B0b & 310.6 & -19.7 & 3      & 0.07      &    1.08 & 1.40 &  0.14 &        1.98 \\
2019/07/07 & second half  &     05:03--05:06              & B0c & 310.2 & -18.0 & 2      & 0.00      &    1.08 & 1.54 &  0.24 &        3.26 \\
2019/07/08 & second half  &     05:12--10:31              & B0a & 308.8 & -19.2 & 121    & 11.99     &    1.20 & 1.14 &  0.30 &        2.37 \\
2019/07/08 & second half  &     04:55--05:01,09:59-10:05  & B0b & 310.6 & -19.7 & 6      & 0.61      &    1.34 & 1.42 &  0.26 &        2.21 \\
2019/07/08 & second half  &     05:04--05:10,10:07-10:12  & B0c & 310.2 & -18.0 & 6      & 0.35      &    1.38 & 1.36 &  0.33 &        2.41 \\
2019/08/27 & full night   &     23:35--23:40,04:37--04:42 & B0a & 307.8 & -19.4 & 6      & 5.12      &    1.27 & 1.05 &  0.11 &        0.00 \\
2019/08/27 & full night   &     23:15--23:32,04:29--04:34 & B0b & 309.6 & -19.9 & 11     & 5.20      &    1.42 & 1.28 &  0.93 &        0.19 \\
2019/08/27 & full night   &     23:42--04:26              & B0c & 309.2 & -18.2 & 115    & 98.31     &    1.11 & 1.06 &  0.00 &        0.00 \\
2019/08/27 & full night   &     04:57--04:59,09:27--09:32 & B1a & 351.9 & -5.0  & 6      & 5.60      &    1.51 & 1.00 &  0.15 &        0.00 \\
2019/08/27 & full night   &     04:46--04:52,09:20--09:25 & B1b & 353.6 & -5.3  & 6      & 5.06      &    1.44 & 1.02 &  0.12 &        0.00 \\
2019/08/27 & full night   &     05:02--09:17              & B1c & 352.9 & -3.6  & 103    & 62.55     &    1.27 & 1.17 &  0.06 &        0.06 \\
2019/08/28 & full night   &     23:43--04:27              & B0a & 307.8 & -19.4 & 115    & 135.40    &    1.10 & 0.95 &  0.00 &        0.00 \\
2019/08/28 & full night   &     23:35--23:40,04:39--04:44 & B0b & 309.6 & -19.9 & 6      & 5.96      &    1.27 & 1.00 &  0.00 &        0.00 \\
2019/08/28 & full night   &     23:14--23:33,04:29--04:34 & B0c & 309.2 & -18.2 & 9      & 5.96      &    1.38 & 1.02 &  0.81 &        0.00 \\
2019/08/28 & full night   &     05:03--09:15              & B1a & 351.9 & -5.0  & 102    & 53.31     &    1.27 & 1.48 &  0.00 &        0.00 \\
2019/08/28 & full night   &     04:45--05:01,09:28--09:33 & B1b & 353.6 & -5.3  & 6      & 4.06      &    1.49 & 1.28 &  0.01 &        0.00 \\
2019/08/28 & full night   &     04:48--04:53,09:18--09:25 & B1c & 352.9 & -3.6  & 7      & 4.67      &    1.55 & 1.28 &  0.05 &        0.00 \\
2019/08/29 & full night   &     23:17--23:34,04:28--04:35 & B0a & 307.8 & -19.4 & 12     & 6.60      &    1.33 & 1.15 &  0.76 &        0.00 \\
2019/08/29 & full night   &     23:45--04:26              & B0b & 309.6 & -19.9 & 114    & 121.76    &    1.10 & 1.00 &  0.00 &        0.00 \\
2019/08/29 & full night   &     23:36--23:41,04:38--04:43 & B0c & 309.2 & -18.2 & 6      & 5.65      &    1.27 & 1.01 &  0.00 &        0.00 \\
2019/08/29 & full night   &     04:47--04:54,09:18--09:25 & B1a & 351.9 & -5.0  & 8      & 6.57      &    1.51 & 1.17 &  0.00 &        0.00 \\
2019/08/29 & full night   &     05:04--09:15              & B1b & 353.6 & -5.3  & 101    & 113.41    &    1.25 & 0.98 &  0.00 &        0.00 \\
2019/08/29 & full night   &     04:56--05:01,09:30--09:33 & B1c & 352.9 & -3.6  & 6      & 5.33      &    1.56 & 1.12 &  0.00 &        0.00 \\
2019/09/26 & first half   &     00:19--04:16              & B1a & 351.4 & -5.2  & 97     & 57.57     &    1.26 & 1.25 &  0.00 &        0.00 \\
2019/09/26 & first half   &     00:04--00:09,04:23--04:28 & B1b & 353.1 & -5.5  & 6      & 1.40      &    1.52 & 1.88 &  0.22 &        0.00 \\
2019/09/26 & first half   &     00:11--00:16,04:18--04:21 & B1c & 352.4 & -3.8  & 5      & 1.43      &    1.57 & 1.62 &  0.21 &        0.00 \\
2019/09/27 & first half   &     00:01--00:06,04:17--04:22 & B1a & 351.4 & -5.2  & 6      & 1.74      &    1.49 & 1.77 &  0.37 &        0.00 \\
2019/09/27 & first half   &     00:16--04:14              & B1b & 353.1 & -5.5  & 97     & 48.07     &    1.28 & 1.28 &  0.22 &        0.00 \\
2019/09/27 & first half   &     00:08--00:13,04:24--04:29 & B1c & 352.4 & -3.8  & 6      & 1.86      &    1.50 & 1.69 &  0.38 &        0.00 \\
2019/09/28 & first half   &     00:02--00:07,04:17--04:22 & B1a & 351.4 & -5.2  & 6      & 5.25      &    1.46 & 1.07 &  0.10 &        0.00 \\
2019/09/28 & first half   &     00:10--00:15,04:24--04:29 & B1b & 353.1 & -5.5  & 6      & 5.14      &    1.46 & 1.10 &  0.08 &        0.00 \\
2019/09/28 & first half   &     00:18--04:13              & B1c & 352.4 & -3.8  & 96     & 84.00     &    1.28 & 1.06 &  0.00 &        0.00 \\
2020/04/18 & second half  &     no data collected         & --- & ---   & ---   & ---    & ---       &    ---  & ---  &  ---  &        ---  \\
2020/04/19 & second half  &     no data collected         & --- & ---   & ---   & ---    & ---       &    ---  & ---  &  ---  &        ---  \\
2020/04/20 & second half  &     no data collected         & --- & ---   & ---   & ---    & ---       &    ---  & ---  &  ---  &        ---  \\
2020/04/21 & second half  &     no data collected         & --- & ---   & ---   & ---    & ---       &    ---  & ---  &  ---  &        ---  \\
2020/04/22 & second half  &     no data collected         & --- & ---   & ---   & ---    & ---       &    ---  & ---  &  ---  &        ---  \\
2020/04/23 & second half  &     no data collected         & --- & ---   & ---   & ---    & ---       &    ---  & ---  &  ---  &        ---  \\
2020/05/17 & second half  &     no data collected         & --- & ---   & ---   & ---    & ---       &    ---  & ---  &  ---  &        ---  \\
2020/05/18 & second half  &     no data collected         & --- & ---   & ---   & ---    & ---       &    ---  & ---  &  ---  &        ---  \\
2020/05/19 & second half  &     no data collected         & --- & ---   & ---   & ---    & ---       &    ---  & ---  &  ---  &        ---  \\
2020/05/20 & second half  &     no data collected         & --- & ---   & ---   & ---    & ---       &    ---  & ---  &  ---  &        ---  \\
2020/05/21 & second half  &     no data collected         & --- & ---   & ---   & ---    & ---       &    ---  & ---  &  ---  &        ---  \\
2020/05/22 & second half  &     no data collected         & --- & ---   & ---   & ---    & ---       &    ---  & ---  &  ---  &        ---  \\
2020/10/15 & first half   &     23:58--04:11              & B1b & 352.9 & -5.6  & 103    & 59.94     &    1.17 & 1.38 &  0.00 &        0.00 \\
2020/10/15 & first half   &     23:53--23:56,04:14--04:18 & B1c & 352.2 & -3.9  & 6      & 3.38      &    1.34 & 1.16 &  0.39 &        0.00 \\
2020/10/15 & first half   &     23:43--23:48,04:21--04:26 & B1e & 353.9 & -4.2  & 6      & 2.57      &    1.37 & 1.28 &  0.99 &        0.00 \\
2020/10/16 & first half   &     23:43--23:48,04:12--04:17 & B1b & 352.9 & -5.6  & 6      & 2.84      &    1.33 & 1.08 &  1.17 &        0.00 \\
2020/10/16 & first half   &     00:13--04:09              & B1c & 352.2 & -3.9  & 96     & 84.45     &    1.17 & 1.07 &  0.00 &        0.00 \\
2020/10/16 & first half   &     23:58--00:03              & B1d & 354.6 & -5.9  & 3      & 1.37      &    1.40 & 1.16 &  0.41 &        0.00 \\
2020/10/16 & first half   &     23:51--23:56,04:19--04:24 & B1e & 353.9 & -4.2  & 6      & 3.32      &    1.34 & 1.13 &  0.53 &        0.00 \\
2020/10/16 & first half   &     00:06--00:11              & B1f & 353.2 & -2.5  & 3      & 1.87      &    1.39 & 1.11 &  0.15 &        0.00 \\
2020/10/17 & first half   &     23:45--23:50,04:14--04:19 & B1b & 352.9 & -5.6  & 6      & 3.48      &    1.32 & 1.17 &  1.02 &        0.00 \\
2020/10/17 & first half   &     23:52--23:58,04:21--04:26 & B1c & 352.2 & -3.9  & 6      & 3.26      &    1.34 & 1.16 &  0.40 &        0.00 \\
2020/10/17 & first half   &     00:00--00:05              & B1d & 354.6 & -5.9  & 3      & 1.16      &    1.37 & 1.25 &  0.41 &        0.00 \\
2020/10/17 & first half   &     00:15--04:12              & B1e & 353.9 & -4.2  & 95     & 88.66     &    1.17 & 1.03 &  0.00 &        0.00 \\
2020/10/17 & first half   &     00:08--00:13              & B1f & 353.2 & -2.5  & 3      & 1.50      &    1.37 & 1.23 &  0.16 &        0.00 \\
2020/10/18 & first half   &     00:20--04:04              & B1a & 351.1 & -5.3  & 91     & 104.06    &    1.15 & 0.89 &  0.00 &        0.00 \\
2020/10/18 & first half   &     00:11--00:16              & B1b & 352.9 & -5.6  & 3      & 3.14      &    1.29 & 0.84 &  0.18 &        0.00 \\
2020/10/18 & first half   &     23:48--23:53,04:06--04:11 & B1d & 354.6 & -5.9  & 6      & 2.91      &    1.31 & 1.06 &  0.83 &        0.00 \\
2020/10/18 & first half   &     23:56--00:01,04:14--04:18 & B1e & 353.9 & -4.2  & 6      & 3.41      &    1.32 & 1.06 &  0.38 &        0.00 \\
2020/10/18 & first half   &     00:03--00:08,04:21--04:26 & B1f & 353.2 & -2.5  & 6      & 4.32      &    1.33 & 1.02 &  0.18 &        0.00 \\
2020/10/19 & first half   &     23:49--23:54,04:09--04:17 & B1a & 351.1 & -5.3  & 7      & 1.76      &    1.30 & 1.56 &  0.74 &        0.00 \\
2020/10/19 & first half   &     23:56--00:01,04:19--04:24 & B1c & 352.2 & -3.9  & 6      & 2.32      &    1.32 & 1.35 &  0.45 &        0.00 \\
2020/10/19 & first half   &     00:04--04:07              & B1d & 354.6 & -5.9  & 99     & 51.48     &    1.16 & 1.24 &  0.16 &        0.00 \\
2020/10/20 & first half   &     23:47--23:59,04:11--04:21 & B1b & 352.9 & -5.6  & 6      & 2.61      &    1.30 & 1.27 &  0.90 &        0.00 \\
2020/10/20 & first half   &     23:52--00:02,04:14--04:24 & B1e & 353.9 & -4.2  & 6      & 2.87      &    1.32 & 1.24 &  0.69 &        0.00 \\
2020/10/20 & first half   &     00:04--04:09              & B1f & 353.2 & -2.5  & 98     & 64.84     &    1.19 & 1.10 &  0.20 &        0.00 \\
2020/10/21 & first half   &     23:49--23:59,04:11--04:22 & B1a & 351.1 & -5.3  & 7      & 3.30      &    1.30 & 1.04 &  0.86 &        0.00 \\
2020/10/21 & first half   &     23:51--00:01,04:14--04:24 & B1c & 352.2 & -3.9  & 6      & 2.61      &    1.32 & 1.10 &  0.82 &        0.00 \\
2020/10/21 & first half   &     00:04--04:08              & B1d & 354.6 & -5.9  & 99     & 75.08     &    1.15 & 0.96 &  0.27 &        0.00 \\
2021/05/03 & first half   &     00:28--00:33,04:34--04:39 & A0f & 217.7 & -11.5 & 6      & 1.04      &    1.46 & 2.09 &  0.18 &        0.00 \\
2021/05/03 & first half   &     00:14--00:26,04:24--04:31 & A0i & 218.9 & -12.8 & 8      & 1.53      &    1.53 & 2.03 &  0.18 &        0.15 \\
2021/05/03 & first half   &     00:36--04:22              & A0j & 219.5 & -11.2 & 92     & 29.16     &    1.30 & 1.83 &  0.00 &        0.05 \\
2021/05/04 & first half   &     02:29--04:38              & A0f & 217.7 & -11.5 & 53     & 26.49     &    1.09 & 1.48 &  0.00 &        0.00 \\
2021/05/05 & no time scheduled & no data collected         & --- & ---   & ---   & ---    & ---       &    ---  & ---  &  ---  &        ---  \\
2021/05/06 & full night   &     00:20--01:52              & A0i & 218.9 & -12.8 & 26     & 0.08      &    1.51 & 1.68 &  0.47 &        3.12 \\
2021/05/06 & full night   &     02:54--10:30              & A1j & 256.7 & -19.3 & 179    & 22.10     &    1.19 & 1.69 &  0.39 &        1.88 \\
2021/05/07 & full night   &     23:53--00:00,05:18--05:24 & A0e & 217.1 & -13.1 & 6      & 0.29      &    1.55 & 1.90 &  0.34 &        2.01 \\
2021/05/07 & full night   &     00:04--05:05              & A0i & 218.9 & -12.8 & 121    & 43.06     &    1.26 & 1.54 &  0.05 &        0.26 \\
2021/05/07 & full night   &     23:46--23:51,05:09--05:14 & A0j & 219.5 & -11.2 & 6      & 0.42      &    1.72 & 1.68 &  0.33 &        1.14 \\
2021/05/07 & full night   &     05:36--05:41              & A1e & 254.7 & -21.9 & 2      & 0.00      &    1.04 & 1.31 &  0.33 &        5.57 \\
2021/05/07 & full night   &     05:45--10:28              & A1f & 255.0 & -20.1 & 88     & 0.39      &    1.17 & 1.35 &  0.65 &        3.87 \\
2021/05/07 & full night   &     05:28--05:34              & A1i & 256.4 & -21.1 & 3      & 0.00      &    1.06 & 1.42 &  0.42 &        3.41 \\
2021/05/08 & no time scheduled &    no data collected         & --- & ---   & ---   & ---    & ---       &    ---  & ---  &  ---  &        ---  \\
2021/05/09 & full night   &     23:55--02:39              & A0g & 217.8 & -16.1 & 66     & 5.03      &    1.37 & 1.33 &  0.40 &        2.07 \\
2021/05/09 & full night   &     06:21--06:24,10:20--10:25 & A1e & 254.7 & -21.9 & 5      & 0.33      &    1.39 & 1.07 &  1.52 &        0.86 \\
2021/05/09 & full night   &     06:26--10:10              & A1f & 255.0 & -20.1 & 68     & 5.61      &    1.18 & 1.06 &  0.60 &        1.83 \\
2021/05/09 & full night   &     06:11--06:16,10:13-10:18  & A1i & 256.4 & -21.1 & 6      & 0.32      &    1.28 & 1.14 &  0.86 &        1.35 \\
2021/05/10 & full night   &     23:59--00:10,04:01--04:06 & A0a & 214.1 & -12.1 & 7      & 4.68      &    1.41 & 1.18 &  0.03 &        0.00 \\
2021/05/10 & full night   &     00:13--00:18,04:08--04:13 & A0b & 215.3 & -13.4 & 6      & 4.56      &    1.34 & 1.10 &  0.02 &        0.00 \\
2021/05/10 & full night   &     00:20--03:53              & A0c & 215.9 & -11.8 & 73     & 58.44     &    1.21 & 1.09 &  0.00 &        0.00 \\
2021/05/10 & full night   &     04:17--04:22,10:17--10:22 & A1c & 253.2 & -20.8 & 6      & 2.68      &    1.43 & 1.42 &  0.73 &        0.05 \\
2021/05/10 & full night   &     04:35--06:59              & A1f & 255.0 & -20.1 & 40     & 25.38     &    1.04 & 1.15 &  0.11 &        0.14 \\
2021/05/10 & full night   &     10:25--10:30              & A1h & 256.2 & -22.9 & 3      & 0.11      &    1.64 & 1.53 &  2.57 &        0.00 \\
2021/05/10 & full night   &     08:06--10:15              & A1i & 256.4 & -21.1 & 45     & 12.95     &    1.30 & 1.47 &  0.44 &        0.10 \\
2021/05/11 & no time scheduled &    no data collected         & --- & ---   & ---   & ---    & ---       &    ---  & ---  &  ---  &        ---  \\
2021/05/12 & full night   &     23:01--23:03,01:05--01:10 & A0a & 214.1 & -12.1 & 5      & 1.30      &    1.74 & 1.54 &  0.48 &        0.14 \\
2021/05/12 & full night   &     01:12--01:15              & A0c & 215.9 & -11.8 & 2      & 1.07      &    1.30 & 1.21 &  0.14 &        0.00 \\
2021/05/12 & full night   &     01:20--01:59              & A0e & 217.1 & -13.1 & 17     & 9.22      &    1.21 & 1.06 &  0.13 &        0.35 \\
2021/05/13 & full night   &     23:24--23:29              & A0b & 215.3 & -13.4 & 3      & 1.28      &    1.99 & 1.15 &  0.48 &        0.00 \\
2021/05/13 & full night   &     23:32--23:37              & A0d & 216.6 & -14.8 & 3      & 1.54      &    1.92 & 1.07 &  0.44 &        0.05 \\
2021/05/13 & full night   &     23:39--03:45              & A0e & 217.1 & -13.1 & 100    & 110.09    &    1.28 & 0.91 &  0.01 &        0.00 \\
2021/05/13 & full night   &     03:49--07:10              & A1c & 253.2 & -20.8 & 82     & 94.55     &    1.06 & 0.87 &  0.03 &        0.00 \\
2021/05/13 & full night   &     07:17--10:30              & A1e & 254.7 & -21.9 & 48     & 28.99     &    1.33 & 1.11 &  0.40 &        0.07 \\
2021/05/14 & no time scheduled &    no data collected         & --- & ---   & ---   & ---    & ---       &    ---  & ---  &  ---  &        ---  \\
2021/05/15 & full night   &     02:15--06:33              & A0h & 218.4 & -14.4 & 101    & 66.56     &    1.11 & 1.19 &  0.01 &        0.12 \\
2021/05/15 & full night   &     06:37--10:31              & A1h & 256.2 & -22.9 & 74     & 27.91     &    1.27 & 1.33 &  0.40 &        0.21 \\
2021/05/16 & full night   &     23:18--03:09              & A0a & 214.1 & -12.1 & 94     & 70.16     &    1.30 & 1.04 &  0.16 &        0.00 \\
2021/05/16 & full night   &     03:28--06:48              & A0b & 215.3 & -13.4 & 79     & 72.86     &    1.19 & 0.99 &  0.00 &        0.00 \\
2021/05/16 & full night   &     22:48--22:53,03:13--03:18 & A0d & 216.6 & -14.8 & 6      & 2.01      &    1.72 & 1.30 &  1.26 &        0.15 \\
2021/05/16 & full night   &     22:56--23:01,03:20--03:25 & A0g & 217.8 & -16.1 & 6      & 2.30      &    1.67 & 1.29 &  0.69 &        0.15 \\
2021/05/16 & full night   &     06:52--10:30              & A1a & 251.5 & -21.4 & 89     & 53.49     &    1.39 & 1.07 &  0.43 &        0.00 \\
2021/05/17 & no time scheduled &    no data collected         & --- & ---   & ---   & ---    & ---       &    ---  & ---  &  ---  &        ---  \\
2021/05/18 & second half  &     04:38--04:43              & A1a & 251.5 & -21.4 & 3      & 0.56      &    1.04 & 2.00 &  0.00 &        0.44 \\
2021/05/18 & second half  &     04:53--04:58              & A1b & 253.0 & -22.6 & 3      & 0.79      &    1.03 & 1.88 &  0.00 &        0.26 \\
2021/05/18 & second half  &     05:23--07:43              & A1d & 254.5 & -23.7 & 57     & 2.96      &    1.03 & 1.74 &  0.19 &        1.58 \\
2021/05/18 & second half  &     05:01--05:05              & A1e & 254.7 & -21.9 & 3      & 0.23      &    1.03 & 2.23 &  0.00 &        0.76 \\
2021/05/18 & second half  &     05:10--05:13              & A1f & 255.0 & -20.1 & 2      & 0.11      &    1.03 & 2.47 &  0.00 &        0.81 \\
2021/05/18 & second half  &     04:46--04:51              & A1h & 256.2 & -22.9 & 3      & 0.45      &    1.05 & 2.32 &  0.00 &        0.29 \\
2021/05/18 & second half  &     05:15--05:20              & A1j & 256.7 & -19.3 & 3      & 0.08      &    1.03 & 2.55 &  0.02 &        1.05 \\
2021/05/19 & second half  &     no data collected         & --- & ---   & ---   & ---    & ---       &    ---  & ---  &  ---  &        ---  \\
2021/09/03 & first half   &     23:25--02:07              & B0b & 309.4 & -19.9 & 66     & 31.86     &    1.13 & 1.44 &  0.00 &        0.08 \\
2021/09/03 & first half   &     02:11--04:40              & B1f & 353.8 & -2.3  & 61     & 24.51     &    1.34 & 1.66 &  0.00 &        0.00 \\
2021/09/04 & first half   &     23:21--02:12              & B0a & 307.6 & -19.4 & 70     & 49.42     &    1.12 & 1.18 &  0.03 &        0.00 \\
2021/09/04 & first half   &     02:16--04:40              & B1e & 354.5 & -3.9  & 59     & 35.91     &    1.30 & 1.33 &  0.00 &        0.00 \\
2021/09/05 & first half   &     23:21--00:14              & B0a & 307.6 & -19.4 & 12     & 6.06      &    1.22 & 1.11 &  0.56 &        0.00 \\
2021/09/05 & first half   &     23:29--02:01              & B0c & 309.0 & -18.3 & 52     & 46.85     &    1.10 & 0.98 &  0.08 &        0.00 \\
2021/09/05 & first half   &     02:05--02:25              & B1e & 354.5 & -3.9  & 6      & 3.14      &    1.60 & 1.26 &  0.08 &        0.00 \\
2021/09/05 & first half   &     02:12--02:32              & B1f & 353.8 & -2.3  & 6      & 3.61      &    1.56 & 1.18 &  0.05 &        0.00 \\
2021/09/05 & first half   &     02:35--04:41              & B1i & 355.5 & -2.5  & 52     & 28.12     &    1.29 & 1.40 &  0.00 &        0.05 \\
2021/09/06 & first half   &     23:22--23:27              & B0c & 309.0 & -18.3 & 3      & 0.47      &    1.32 & 1.25 &  1.49 &        0.00 \\
2021/09/06 & first half   &     23:29--02:18              & B0f & 310.4 & -17.1 & 69     & 41.17     &    1.12 & 1.23 &  0.01 &        0.05 \\
2021/09/06 & first half   &     02:30--04:39              & B1c & 352.8 & -3.7  & 53     & 30.45     &    1.24 & 1.29 &  0.00 &        0.00 \\
2021/09/06 & first half   &     02:22--02:27              & B1i & 355.5 & -2.5  & 3      & 1.17      &    1.55 & 1.43 &  0.11 &        0.08 \\
2021/09/07 & first half   &     23:25--23:30              & B0a & 307.6 & -19.4 & 3      & 0.03      &    1.26 & 5.13 &  1.18 &        0.19 \\
2021/09/07 & first half   &     23:33--02:04              & B0d & 311.3 & -20.4 & 62     & 3.19      &    1.11 & 4.54 &  0.00 &        0.15 \\
2021/09/07 & first half   &     02:08--02:13              & B1c & 352.8 & -3.7  & 3      & 0.03      &    1.53 & 7.94 &  0.00 &        0.12 \\
2021/09/07 & first half   &     02:16--04:37              & B1h & 356.2 & -4.2  & 58     & 2.02      &    1.29 & 5.83 &  0.00 &        0.00 \\
2021/09/08 & first half   &     23:22--23:27              & B0a & 307.6 & -19.4 & 3      & 0.36      &    1.25 & 1.18 &  1.83 &        0.00 \\
2021/09/08 & first half   &     23:29--01:11              & B0d & 311.3 & -20.4 & 42     & 15.13     &    1.14 & 1.37 &  0.32 &        0.16 \\
2021/09/08 & first half   &     01:36--01:41              & B1c & 352.8 & -3.7  & 3      & 1.24      &    1.73 & 1.25 &  0.34 &        0.02 \\
2021/09/08 & first half   &     01:44--04:35              & B1h & 356.2 & -4.2  & 70     & 32.49     &    1.35 & 1.36 &  0.06 &        0.08 \\
2021/09/09 & first half   &     23:35--23:50              & B0a & 307.6 & -19.4 & 4      & 4.40      &    1.18 & 0.83 &  0.19 &        0.00 \\
2021/09/09 & first half   &     00:00--01:37              & B0c & 309.0 & -18.3 & 40     & 50.56     &    1.08 & 0.83 &  0.02 &        0.00 \\
2021/09/09 & first half   &     23:53--23:58              & B0d & 311.3 & -20.4 & 3      & 2.37      &    1.18 & 1.41 &  0.14 &        0.00 \\
2021/09/09 & first half   &     23:38--23:43              & B0f & 310.4 & -17.1 & 3      & 2.98      &    1.24 & 0.85 &  0.21 &        0.00 \\
2021/09/09 & first half   &     01:57--04:36              & B1a & 351.7 & -5.1  & 65     & 63.82     &    1.24 & 1.08 &  0.00 &        0.00 \\
2021/09/09 & first half   &     01:42--01:47              & B1b & 353.5 & -5.4  & 3      & 1.77      &    1.63 & 1.23 &  0.00 &        0.00 \\
2021/09/09 & first half   &     01:50--01:55              & B1e & 354.5 & -3.9  & 3      & 2.30      &    1.64 & 1.08 &  0.00 &        0.00 \\
2021/09/10 & first half   &     23:19--00:59              & B0a & 307.6 & -19.4 & 41     & 3.93      &    1.12 & 1.08 &  1.08 &        1.95 \\
2021/09/10 & first half   &     01:05--04:06              & B1a & 351.7 & -5.1  & 66     & 10.51     &    1.36 & 1.05 &  0.24 &        2.20 \\
2021/09/11 & first half   &     no data collected         & --- & ---   & ---   & ---    & ---       &    ---  & ---  &  ---  &        ---  \\
2021/09/12 & first half   &     23:24--01:14              & B0e & 310.8 & -18.7 & 45     & 14.32     &    1.12 & 1.33 &  0.68 &        0.01 \\
2021/09/12 & first half   &     01:18--04:09              & B1b & 353.5 & -5.4  & 70     & 20.78     &    1.33 & 1.52 &  0.39 &        0.03 \\
2021/09/27 & first half   &     00:16--03:54              & B1d & 354.8 & -5.8  & 89     & 78.75     &    1.30 & 1.03 &  0.00 &        0.00 \\
2021/09/27 & first half   &     00:01--00:06,04:04--04:09 & B1g & 356.5 & -6.0  & 6      & 3.78      &    1.58 & 1.22 &  0.11 &        0.00 \\
2021/09/27 & first half   &     23:53--00:14,03:57--04:02 & B1j & 354.4 & -1.0  & 9      & 4.39      &    1.80 & 1.25 &  0.27 &        0.00 \\
2021/09/28 & first half   &     00:05--00:09,04:00--04:05 & B1d & 354.8 & -5.8  & 6      & 3.81      &    1.49 & 1.30 &  0.13 &        0.24 \\
2021/09/28 & first half   &     00:12--03:57              & B1g & 356.5 & -6.0  & 92     & 62.64     &    1.32 & 1.05 &  0.10 &        0.29 \\
2021/09/28 & first half   &     23:57--00:02,04:07--04:12 & B1j & 354.4 & -1.0  & 6      & 4.15      &    1.64 & 1.19 &  0.19 &        0.03 \\
2021/09/29 & first half   &     no data collected         & --- & ---   & ---   & ---    & ---       &    ---  & ---  &  ---  &        ---  \\
2021/09/30 & first half   &     00:11--00:16,03:58--04:05 & B1d & 354.8 & -5.8  & 7      & 1.58      &    1.37 & 2.03 &  0.22 &        0.05 \\
2021/09/30 & first half   &     00:20--00:24,04:07--04:12 & B1g & 356.5 & -6.0  & 6      & 1.27      &    1.41 & 1.86 &  0.24 &        0.00 \\
2021/09/30 & first half   &     00:27--03:55              & B1j & 354.4 & -1.0  & 85     & 33.11     &    1.31 & 1.46 &  0.22 &        0.00 \\
2021/10/01 & first half   &     00:37--03:56              & B1b & 353.1 & -5.5  & 81     & 84.68     &    1.21 & 0.88 &  0.14 &        0.00 \\
2021/10/01 & first half   &     00:20--00:27,03:58--04:06 & B1f & 353.4 & -2.4  & 8      & 8.55      &    1.40 & 0.86 &  0.29 &        0.00 \\
2021/10/01 & first half   &     00:30--00:35,04:08--04:13 & B1i & 355.1 & -2.7  & 6      & 5.96      &    1.39 & 0.88 &  0.30 &        0.00 \\
2021/10/02 & first half   &     23:58--00:05,03:58--04:06 & B1b & 353.1 & -5.5  & 8      & 5.73      &    1.42 & 1.01 &  0.30 &        0.00 \\
2021/10/02 & first half   &     00:15--03:56              & B1f & 353.4 & -2.4  & 89     & 84.73     &    1.27 & 0.90 &  0.19 &        0.00 \\
2021/10/02 & first half   &     00:08--00:13,04:08--04:13 & B1i & 355.1 & -2.7  & 6      & 4.51      &    1.47 & 0.99 &  0.29 &        0.00 \\
2021/10/03 & first half   &     23:28--23:46,04:13        & B1b & 353.1 & -5.5  & 4      & 1.40      &    1.69 & 1.12 &  1.28 &        0.00 \\
2021/10/03 & first half   &     23:50--23:58              & B1f & 353.4 & -2.4  & 3      & 0.94      &    1.85 & 1.58 &  0.59 &        0.00 \\
2021/10/03 & first half   &     00:02--04:10              & B1i & 355.1 & -2.7  & 100    & 84.57     &    1.30 & 0.99 &  0.10 &        0.00 \\
2021/10/04 & first half   &     23:58--04:08              & B1c & 352.4 & -3.9  & 101    & 81.26     &    1.25 & 1.11 &  0.00 &        0.00 \\
2021/10/04 & first half   &     23:39--23:48              & B1e & 354.1 & -4.1  & 4      & 0.50      &    1.91 & 1.36 &  1.60 &        0.00 \\
2021/10/04 & first half   &     23:50--23:55,04:10--04:15 & B1h & 355.8 & -4.4  & 6      & 5.44      &    1.50 & 1.04 &  0.25 &        0.04 \\
2021/10/05 & first half   &     23:44--23:52,04:02--04:07 & B1c & 352.4 & -3.9  & 7      & 3.70      &    1.49 & 1.28 &  0.68 &        0.00 \\
2021/10/05 & first half   &     23:54--23:59,04:10--04:15 & B1e & 354.1 & -4.1  & 6      & 4.69      &    1.44 & 1.00 &  0.36 &        0.00 \\
2021/10/05 & first half   &     00:02--04:00              & B1h & 355.8 & -4.4  & 97     & 73.04     &    1.27 & 1.02 &  0.17 &        0.01 \\
2021/10/06 & first half   &     23:50-23:55,04:05--04:10  & B1c & 352.4 & -3.9  & 5      & 2.51      &    1.48 & 1.17 &  0.53 &        0.00 \\
2021/10/06 & first half   &     00:05--04:03              & B1e & 354.1 & -4.1  & 97     & 72.75     &    1.24 & 1.09 &  0.07 &        0.00 \\
2021/10/06 & first half   &     23:57--00:02,04:08--04:13 & B1h & 355.8 & -4.4  & 5      & 2.58      &    1.49 & 1.16 &  0.36 &        0.02 \\
2022/05/25 & first half   &     23:04--02:05              & A0a & 213.7 & -12.0 & 74     & 79.04     &    1.27 & 0.84 &  0.21 &        0.00 \\
2022/05/25 & first half   &     22:50--22:59,04:33--04:38 & A0b & 215.0 & -13.3 & 6      & 4.26      &    1.48 & 0.96 &  1.07 &        0.00 \\
2022/05/25 & first half   &     22:52--04:30              & A0f & 217.3 & -11.4 & 61     & 75.58     &    1.12 & 0.77 &  0.04 &       10.87 \\
2022/05/26 & first half   &     22:53--23:23              & A0a & 213.7 & -12.0 & 5      & 1.38      &    1.64 & 1.28 &  1.06 &        0.00 \\
2022/05/26 & first half   &     23:26--02:19              & A0b & 215.0 & -13.3 & 71     & 79.08     &    1.21 & 0.88 &  0.09 &        0.00 \\
2022/05/26 & first half   &     22:51--04:38              & A0c & 215.5 & -11.7 & 61     & 79.44     &    1.14 & 0.86 &  0.03 &        0.00 \\
2022/05/26 & first half   &     22:48--23:18              & A0f & 217.3 & -11.4 & 5      & -0.16     &    1.83 & 1.28 &  1.58 &       19.55 \\
2022/05/27 & first half   &     22:47--23:09              & A0a & 213.7 & -12.0 & 4      & 0.07      &    1.69 & 0.95 &  1.96 &       24.27 \\
2022/05/27 & first half   &     22:49--23:12              & A0b & 215.0 & -13.3 & 4      & 1.33      &    1.68 & 0.92 &  1.66 &        0.00 \\
2022/05/27 & first half   &     22:52--23:14              & A0c & 215.5 & -11.7 & 4      & 1.49      &    1.72 & 0.94 &  1.40 &        0.00 \\
2022/05/27 & first half   &     23:16--02:12              & A0d & 216.2 & -14.6 & 72     & 83.21     &    1.22 & 0.87 &  0.08 &        0.00 \\
2022/05/27 & first half   &     02:14--04:38              & A0e & 216.8 & -13.0 & 59     & 68.77     &    1.08 & 0.86 &  0.05 &        0.00 \\
2022/05/28 & first half   &     no data collected         & --- & ---   & ---   & ---    & ---       &    ---  & ---  &  ---  &        ---  \\
2022/05/29 & no time scheduled &    no data collected         & --- & ---   & ---   & ---    & ---       &    ---  & ---  &  ---  &        ---  \\
2022/05/30 & first half   &     no data collected         & --- & ---   & ---   & ---    & ---       &    ---  & ---  &  ---  &        ---  \\
2022/05/31 & first half   &     no data collected         & --- & ---   & ---   & ---    & ---       &    ---  & ---  &  ---  &        ---  \\
2022/06/01 & first half   &     no data collected         & --- & ---   & ---   & ---    & ---       &    ---  & ---  &  ---  &        ---  \\
2022/06/02 & first half   &     no data collected         & --- & ---   & ---   & ---    & ---       &    ---  & ---  &  ---  &        ---  \\
2022/06/03 & first half   &     no data collected         & --- & ---   & ---   & ---    & ---       &    ---  & ---  &  ---  &        ---  \\
2022/06/04 & first half   &     no data collected         & --- & ---   & ---   & ---    & ---       &    ---  & ---  &  ---  &        ---  \\
2022/06/05 & first half   &     no data collected         & --- & ---   & ---   & ---    & ---       &    ---  & ---  &  ---  &        ---  \\
2022/06/06 & first half   &     no data collected         & --- & ---   & ---   & ---    & ---       &    ---  & ---  &  ---  &        ---  \\
2022/08/21 & second half  &     05:04--07:43              & B1a & 351.9 & -5.0  & 65     & 63.76     &    1.13 & 1.07 &  0.00 &        0.00 \\
2022/08/21 & second half  &     04:49--04:59              & B1d & 355.4 & -5.5  & 3      & 3.18      &    1.19 & 1.04 &  0.00 &        0.00 \\
2022/08/21 & second half  &     07:46--10:17              & B1h & 356.5 & -4.1  & 62     & 21.60     &    1.47 & 1.43 &  0.45 &        0.24 \\
2022/08/21 & second half  &     04:52--05:02              & B1m & 357.5 & -2.7  & 3      & 3.10      &    1.23 & 1.05 &  0.00 &        0.00 \\
2022/08/22 & second half  &     04:50--05:01,10:05--10:10 & B1a & 351.9 & -5.0  & 5      & 2.80      &    1.55 & 1.38 &  0.77 &        0.00 \\
2022/08/22 & second half  &     05:06--07:37              & B1d & 355.4 & -5.5  & 62     & 72.37     &    1.12 & 0.96 &  0.00 &        0.00 \\
2022/08/22 & second half  &     07:41--10:02              & B1g & 357.2 & -5.8  & 58     & 41.13     &    1.39 & 1.19 &  0.01 &        0.00 \\
2022/08/22 & second half  &     04:52--05:03,10:07--10:13 & B1m & 357.5 & -2.7  & 5      & 2.38      &    1.52 & 1.33 &  0.94 &        0.00 \\
2022/08/23 & second half  &     04:51--05:07,10:00--10:08 & B1a & 351.9 & -5.0  & 5      & 2.86      &    1.55 & 1.13 &  0.59 &        0.00 \\
2022/08/23 & second half  &     04:54--05:09,10:03--10:10 & B1d & 355.4 & -5.5  & 5      & 2.73      &    1.49 & 1.13 &  0.73 &        0.03 \\
2022/08/23 & second half  &     04:48--05:04,09:57--10:13 & B1g & 357.2 & -5.8  & 6      & 2.62      &    1.53 & 1.17 &  0.89 &        0.03 \\
2022/08/23 & second half  &     05:12--09:55              & B1m & 357.5 & -2.7  & 115    & 95.50     &    1.28 & 1.09 &  0.00 &        0.00 \\
2022/08/24 & no time scheduled &    no data collected         & --- & ---   & ---   & ---    & ---       &    ---  & ---  &  ---  &        ---  \\
2022/08/25 & second half  &     05:07--05:12              & B1b & 353.7 & -5.3  & 3      & 2.87      &    1.13 & 0.80 &  0.02 &        0.40 \\
2022/08/25 & second half  &     05:14--09:30              & B1e & 354.7 & -3.9  & 91     & 40.35     &    1.23 & 0.92 &  0.00 &        1.57 \\
2022/08/25 & second half  &     05:00--05:04              & B1g & 357.2 & -5.8  & 3      & 3.58      &    1.15 & 0.83 &  0.02 &        0.16 \\
2022/08/25 & second half  &     04:49--04:57              & B1k & 358.9 & -6.0  & 3      & 1.91      &    1.18 & 1.10 &  0.03 &        0.16 \\
2022/08/26 & second half  &     09:47--09:52              & B1b & 353.7 & -5.3  & 3      & 1.64      &    2.00 & 1.07 &  0.38 &        0.00 \\
2022/08/26 & second half  &     09:39--09:44              & B1e & 354.7 & -3.9  & 3      & 2.31      &    1.90 & 0.98 &  0.18 &        0.00 \\
2022/08/26 & second half  &     09:54--09:59              & B1i & 355.8 & -2.4  & 3      & 0.92      &    2.08 & 1.12 &  0.93 &        0.00 \\
2022/08/26 & second half  &     07:03--09:37              & B1k & 358.9 & -6.0  & 63     & 87.65     &    1.30 & 0.85 &  0.00 &        0.00 \\
2022/08/26 & second half  &     04:35--07:01              & B1l & 358.2 & -4.4  & 60     & 94.09     &    1.14 & 0.85 &  0.00 &        0.00 \\
2022/08/27 & second half  &     04:45--07:12              & B1b & 353.7 & -5.3  & 60     & 60.63     &    1.12 & 0.96 &  0.00 &        0.00 \\
2022/08/27 & second half  &     07:14--09:53              & B1i & 355.8 & -2.4  & 65     & 52.04     &    1.49 & 0.96 &  0.25 &        0.00 \\
2022/08/27 & second half  &     10:03--10:08              & B1k & 358.9 & -6.0  & 3      & 0.24      &    1.96 & 1.11 &  2.46 &        0.00 \\
2022/08/27 & second half  &     09:55--10:00              & B1l & 358.2 & -4.4  & 3      & 0.63      &    1.96 & 1.10 &  1.37 &        0.00 \\
\enddata
\end{deluxetable}
\end{longrotatetable}

\begin{deluxetable}{|l|l|l|}
\tablewidth{6in}
\label{afields}
\tablecaption{Field coordinates for observations in the 2022 A
  Semester (J2000) for 1 June 2022 UT. Field coordinates for
  observation at other epochs were precessed based on the date of the
  year to account for parallax due to the Earth's motion. There were
  no adjustments made for year-to-year orbital motion because the
  increasing size of the fan pattern was intended to account for
  year-to-year orbital motion.}
\tablehead{\colhead{Field Name} & \colhead{RA} & \colhead{Dec}}
\startdata
A0a & 14:14:59.71 & -11:57:03.44 \\
A0b & 14:19:49.09 & -13:17:36.23 \\
A0c & 14:22:11.42 & -11:39:26.43 \\
A0d & 14:24:41.45 & -14:37:49.69 \\
A0e & 14:27:02.70 & -12:59:30.33 \\
A0f & 14:29:22.09 & -11:21:06.26 \\
A0g & 14:29:36.95 & -15:57:42.78 \\
A0h & 14:31:57.11 & -14:19:13.73 \\
A0i & 14:34:15.23 & -12:40:39.61 \\
A0j & 14:36:31.58 & -11:02:01.10 \\
A0k & 14:34:35.73 & -17:17:14.45 \\
A0l & 14:36:54.80 & -15:38:35.57 \\
A0m & 14:39:11.66 & -13:59:51.30 \\
A0n & 14:41:26.57 & -12:21:02.30 \\
A0o & 14:43:39.79 & -10:42:09.21 \\
A1a & 16:44:20.49 & -21:23:01.75 \\
A1b & 16:50:10.37 & -22:30:03.56 \\
A1c & 16:51:16.46 & -20:42:20.34 \\
A1d & 16:56:05.70 & -23:36:19.26 \\
A1e & 16:57:07.87 & -21:48:27.59 \\
A1f & 16:58:08.51 & -20:00:34.51 \\
A1g & 17:02:06.59 & -24:41:47.35 \\
A1h & 17:03:04.71 & -22:53:47.66 \\
A1i & 17:04:01.31 & -21:05:46.68 \\
A1j & 17:04:56.55 & -19:17:44.56 \\
A1k & 17:08:13.16 & -25:46:26.34 \\
A1l & 17:09:07.07 & -23:58:19.08 \\
A1m & 17:09:59.48 & -22:10:10.67 \\
A1n & 17:10:50.57 & -20:22:01.24 \\
A1o & 17:11:40.48 & -18:33:50.91 \\
\enddata

\end{deluxetable}

\begin{deluxetable}{|l|l|l|}
\tablewidth{6in}
\label{bfields}
\tablecaption{Field coordinates for the 2022 B Semester (J2000) for 25
  Aug 2022 UT. Field coordinates for observation at other epochs were
  precessed based on the date of the year to account for parallax due
  to the Earth's motion. There were no adjustments made for
  year-to-year orbital motion because the increasing size of the fan
  pattern was intended to account for year-to-year orbital motion.}
\tablehead{\colhead{Field Name} & \colhead{RA} & \colhead{Dec}}
\startdata
B0a & 20:31:30.33 & -19:23:23.82 \\
B0b & 20:38:47.25 & -19:53:11.52 \\
B0c & 20:37:06.89 & -18:11:47.89 \\
B0d & 20:46:06.72 & -20:21:55.07 \\
B0e & 20:44:21.65 & -18:40:46.41 \\
B0f & 20:42:38.65 & -16:59:34.17 \\
B0g & 20:53:28.60 & -20:49:35.23 \\
B0h & 20:51:38.85 & -19:08:42.02 \\
B0i & 20:49:51.31 & -17:27:44.81 \\
B0j & 20:48:05.73 & -15:46:44.07 \\
B0k & 21:00:52.74 & -21:16:12.90 \\
B0l & 20:58:58.34 & -19:35:35.57 \\
B0m & 20:57:06.29 & -17:54:53.82 \\
B0n & 20:55:16.36 & -16:14:08.14 \\
B0o & 20:53:28.28 & -14:33:18.99 \\
B1a & 23:27:45.53 & -05:00:35.01 \\
B1b & 23:34:43.47 & -05:16:15.90 \\
B1c & 23:31:55.69 & -03:35:48.69 \\
B1d & 23:41:41.65 & -05:31:42.47 \\
B1e & 23:38:52.92 & -03:51:20.81 \\
B1f & 23:36:04.86 & -02:10:57.08 \\
B1g & 23:48:39.98 & -05:46:56.96 \\
B1h & 23:45:50.41 & -04:06:40.19 \\
B1i & 23:43:01.55 & -02:26:21.18 \\
B1j & 23:40:13.11 & -00:46:00.86 \\
B1k & 23:55:38.35 & -06:02:01.66 \\
B1l & 23:52:48.05 & -04:21:49.08 \\
B1m & 23:49:58.50 & -02:41:34.11 \\
B1n & 23:47:09.42 & -01:01:17.68 \\
B1o & 23:44:20.52 & +00:38:59.32 \\
\enddata

\end{deluxetable}

\subsection{Dithering}

We have considered dithering to remove object losses due to the chip
gaps, however we found that this increases efficiency only marginally
at best. Even if an object is in a chip gap in a single epoch, it is
very unlikely to be in a chip gap in the second epoch. The chip gaps
range from 153 pixels (short edge) to 201 pixels (long edge) for the
chips which are 2048 x 4096 pixels. Thus, the gaps cover about
201x4096 + 153x2048 pixels, about 1 million pixels, which is about
12\% of the field. The odds of an object falling in a gap twice in 2
visits is thus about 1.4\%. With a $\sim 5000$ object expected sample,
on average we would expect 71 objects to fall into a gap twice in 2
years. We consider this acceptable loss since dithering would
significantly complicate image processing and asteroid identification
as we expect that pairwise image subtraction may be optimal for
asteroid identification. Thus no dithering was performed during each
field's $\sim 4$ hour observational sequence.

\section{Summary}

We summarize our survey's observational design as follows:

\begin{itemize}

\item We have constructed a survey that covers between 10 and 45
  square degrees of sky to about magnitude $m_R \sim 26.2$ and will
  measure orbital information for distant objects over a timebase of a
  few years. This will lead to the discovery of a few thousand TNOs
  and a determination of about a thousand of their orbits, surpassing
  all prior surveys in terms of object numbers.

\item We will use the CTIO DECam mounted at prime focus of the Blanco
  4m telescope in Cerro Tololo, Chile as it is one of the most
  powerful survey instruments available to the astronomical community.

\item We have aimed to collect about 120 second exposures for about 4
  hours for every field on each half-night of observations. This
  should allow us to achieve depths of about $m_R \sim 26.2$ for
  typical seeing conditions at Cerro Tololo.

\item Our overall cadence of observations is to collect both
  quadrature and opposition observations the first year, and
  thereafter only opposition observations in subsequent years. This
  achieves an astrometric uncertainty of about $\sim 30$ arc-seconds
  or better for most objects for a few years after the survey.

\item We have chosen our primary filter to be the $V\!R$ filter, which
  encompasses the bandpasses of both the traditional Johnson
  Kron-Cousins $V$ and $R$ filters and has a high efficiency from
  about 470 nm to 760 nm. This choice was intended to maximize
  discovery rates as the bandpass is a factor $\sim 2$ superior in
  flux sensitivity to all other filters in common usage with DECam.

\item We have planned for the use of a single additional filter, the
  $i$-band filter, which would allow colors to be measured for TNOs
  discovered purely in the $V\!R$ filter. This appears to be the
  optimal filter configuration for both discovery and color
  measurements given that typical TNO colors range from neutral
  reflectance to some of the reddest objects in the solar system.

\item Our ``fan''-shaped field pattern allows us to track objects'
  sky-plane orbital motion over the nominal 3 year course of the
  survey yielding significant information about the TNO populations'
  orbital characteristics. Although initially designed for 3 years (a
  total of 10 fields), the ``fan'' shape was extended in the last year
  of the survey to account for observations lost during operational
  and national closures due to the COVID-19 pandemic.

\item The actual fields we observed each night were chosen to maximize
  observability for the nights and half-nights we were allocated on
  the first year of the project and to avoid bright field stars. This
  was adjusted for parallactic motion of objects each year based on
  the date of the year of observation.

\end{itemize}

Our final survey will be the largest of its kind ever conducted,
discovering some $\sim 3,500$ TNOs and providing orbital information
for at least one thousand TNOs. In addition, it will exceed all future
predictions for the LSST in terms of depth and should allow a useful
comparison to the LSST in terms of object numbers and TNO properties.

\clearpage

\begin{acknowledgments}

This work is based in part on observations at Cerro Tololo
Inter-American Observatory at NSF’s NOIRLab (NOIRLab Prop. ID
2019A-0337; PI: D. Trilling), which is managed by the Association of
Universities for Research in Astronomy (AURA) under a cooperative
agreement with the National Science Foundation.

This work is supported by the National Aeronautics and Space
Administration under grant No.\ NNX17AF21G issued through the SSO
Planetary Astronomy Program and by the National Science Foundation
under grants No.\ AST-2009096 and AST-1409547. This research was
supported in part through computational resources and services
provided by Advanced Research Computing at the University of Michigan,
Ann Arbor. This work used the Extreme Science and Engineering
Discovery Environment \citep[XSEDE; ][]{XSEDE}, which is supported by
National Science Foundation grant number ACI-1548562. This work used
the XSEDE Bridges GPU and Bridges-2 GPU-AI at the  Pittsburgh
Supercomputing Center through allocation TG-AST200009.

H. Smotherman acknowledges support by NASA under grant
No.\ 80NSSC21K1528 (FINESST). H. Smotherman, M. Juri\'{c} and
P. Bernardinelli acknowledge the support from the University of
Washington College of Arts and Sciences, Department of Astronomy, and
the DiRAC Institute. The DiRAC Institute is supported through generous
gifts from the Charles and Lisa Simonyi Fund for Arts and Sciences and
the Washington Research Foundation. M. Juri\'{c} wishes to acknowledge
the support of the Washington Research Foundation Data Science Term
Chair fund, and the University of Washington Provost’s Initiative in
Data-Intensive Discovery. 

This project used data obtained with the Dark Energy Camera (DECam),
which was constructed by the Dark Energy Survey (DES)
collaboration. Funding for the DES Projects has been provided by the
US Department of Energy, the US National Science Foundation, the
Ministry of Science and Education of Spain, the Science and Technology
Facilities Council of the United Kingdom, the Higher Education Funding
Council for England, the National Center for Supercomputing
Applications at the University of Illinois at Urbana-Champaign, the
Kavli Institute for Cosmological Physics at the University of Chicago,
Center for Cosmology and Astro-Particle Physics at the Ohio State
University, the Mitchell Institute for Fundamental Physics and
Astronomy at Texas A\&M University, Financiadora de Estudos e
Projetos, Funda\c{c}\~{a}o Carlos Chagas Filho de Amparo \'{a}
Pesquisa do Estado do Rio de Janeiro, Conselho Nacional de
Desenvolvimento Cient\'{i}fico e Tecnol\'{o}gico and the
Minist\'{e}rio da Ci\^{e}ncia, Tecnologia e Inova\c{c}\~{a}o, the
Deutsche Forschungsgemeinschaft and the Collaborating Institutions in
the Dark Energy Survey.

The Collaborating Institutions are Argonne National Laboratory, the
University of California at Santa Cruz, the University of Cambridge,
Centro de Investigaciones En\'{e}rgeticas, Medioambientales y
Tecnol\'{o}gicas–Madrid, the University of Chicago, University College
London, the DES-Brazil Consortium, the University of Edinburgh, the
Eidgen\"{o}ssische Technische Hochschule (ETH) Z\"{u}rich, Fermi
National Accelerator Laboratory, the University of Illinois at
Urbana-Champaign, the Institut de Ci\`{e}ncies de l’Espai (IEEC/CSIC),
the Institut de Física d’Altes Energies, Lawrence Berkeley National
Laboratory, the Ludwig-Maximilians Universit\"{a}t M\"{u}nchen and the
associated Excellence Cluster Universe, the University of Michigan,
NSF’s NOIRLab, the University of Nottingham, the Ohio State
University, the OzDES Membership Consortium, the University of
Pennsylvania, the University of Portsmouth, SLAC National Accelerator
Laboratory, Stanford University, the University of Sussex, and Texas
A\&M University.

\end{acknowledgments}

\bibliography{myrefs}

\end{document}